\documentclass[twocolumn]{aastex631}
\usepackage{ulem}

\newcommand{\kms}{km\,s$^{-1}$}

\submitjournal{ApJ}

\shortauthors{Yang et al.}

\begin{document}

\title{Maser Investigation toward Off-Plane Stars (MIOPS): \\detection of SiO masers in the Galactic thick disk and halo}

\correspondingauthor{Yuanwei Wu, Wenjin Yang}
\email{yuanwei.wu@ntsc.ac.cn, wjyang@nju.edu.cn}

\author[0000-0002-3599-6608]{Wenjin Yang}
\affiliation{School of Astronomy \& Space Science, Nanjing University, 163 Xianlin Avenue, Nanjing 210023, People's Republic of China}
\affiliation{Max-Plank-Institut f$\ddot{u}$r Radioastronomie, auf dem H$\ddot{u}$gel 69, 53121 Bonn, Germany}

\author[0000-0003-3188-5983]{Yuanwei Wu}
\affiliation{National Time Service Center, Chinese Academy of Sciences, Xi'an 710600, People's Republic of China}

\author[0000-0002-3866-414X]{Yan Gong}
\affiliation{Max-Plank-Institut f$\ddot{u}$r Radioastronomie, auf dem H$\ddot{u}$gel 69, 53121 Bonn, Germany}

\author[0000-0002-7234-0632]{Nicolas Mauron}
\affiliation{Universit$\acute{e}$ de Montpellier, Laboratoire Univers et Particules de Montpellier CNRS-IN2P3/UM, Place Bataillon, 34095 Montpellier, France}

\author[0000-0003-1353-9040]{Bo Zhang}
\affiliation{Shanghai Astronomical Observatory, Chinese Academy of Sciences, Shanghai 200030, People's Republic of China}

\author[0000-0001-6459-0669]{Karl M. Menten}
\affiliation{Max-Plank-Institut f$\ddot{u}$r Radioastronomie, auf dem H$\ddot{u}$gel 69, 53121 Bonn, Germany}

\author[0000-0001-7573-0145]{Xiaofeng Mai}
\affiliation{Shanghai Astronomical Observatory, Chinese Academy of Sciences, Shanghai 200030, People's Republic of China}
\affiliation{School of Astronomy and Space Sciences, University of Chinese Academy of Sciences, No. 19A Yuquan Road, Beijing 100049, People's Republic of China}

\author{Dejian Liu}
\affiliation{Purple Mountain Observatory, Chinese Academy of Sciences, Nanjing 210023, People's Republic of China}
\affiliation{School of Astronomy and Space Science, University of Science and Technology of China, Hefei 230026, People's Republic of China}

\author[0000-0003-3520-6191]{Juan Li}
\affiliation{Shanghai Astronomical Observatory, Chinese Academy of Sciences, Shanghai 200030, People's Republic of China}

\author[0000-0002-3338-8465]{Jingjing Li}
\affiliation{Purple Mountain Observatory, Chinese Academy of Sciences, Nanjing 210023, People's Republic of China}



\begin{abstract}
Studying stars that are located off the Galactic plane is important for understanding the formation history of the Milky Way. We searched for SiO masers toward off-plane O-rich asymptotic giant branch (AGB) stars from the catalog presented by \cite{Mauron2019} in order to shed light on the origin of these objects. A total of 102 stars were observed in the SiO $J$=1--0, $v=1$ and 2 transitions with the Effelsberg-100 m and Tianma-65 m telescopes. SiO masers were discovered in eight stars, all first detections. 
The measured maser velocities allow the first estimates of the host AGB stars' radial velocities. 
We find that the radial velocities of three stars (namely G068.881$-$24.615, G070.384$-$24.886, and G084.453$-$21.863) significantly deviate from the values expected from  Galactic circular motion. The updated distances and 3D motions indicate that G068.881$-$24.615 is likely located in the Galactic halo, while G160.648$-$08.846 is probably located in the Galactic thin disk, and the other six stars are probably part of the Galactic thick disk. 
\end{abstract}



\section{Introduction} \label{sec:intro}

Asymptotic Giant Branch (AGB) stars are in the late stages of stellar evolution and, due to their mass loss, are dominating contributors to the cosmic gas/dust cycle \citep[e.g.,][]{2018A&ARv..26....1H}. With typical ages of a few Gyr \citep[e.g.,][]{1999MNRAS.310..629S,2018A&ARv..26....1H}, AGB stars are found 
in all structures of the Milky Way\footnote{Here is a brief description of each Galactic component based on the reviews \citep{2007gitu.book.....S,2016ARA&A..54..529B,2020ARA&A..58..205H}. The thin disk is defined by its flat, disk-like shape and hosts young stars, gas, and dust, all of which generally follow near-circular orbits around the Galactic center. Compared to the thin disk, the thick disk has a higher scale height, and consists of older and lower metallicity stars with more eccentric and inclined orbits. The bulge, located around the Galactic center, harbors a dense mix of old and intermediate-age stars within its spherical structure. Lastly, the halo, the outermost and least luminous component of the Milky Way, comprises very old stars, globular clusters, and dark matter, featuring stars with highly elliptical and randomly oriented orbits.} including the Galactic thin disk, thick disk, bulge, halo, and even the tidal stellar streams produced by the interaction of the Milky Way with its satellites \citep[e.g.,][]{2019ApJ...874..138L,2021ApJS..256...43S,2023ApJS..264...20I}. The 3 dimensional (3D) motions of AGB stars located off the Galactic disk 
can provide a crucial clue on their nature, which potentially 
sheds light on the formation history of our Galaxy. However, so far such 3D motions have only been determined for a few Galactic off-plane AGB stars 
and dedicated studies are needed to determine these objects' motions and origins.


Accurate measurements of 3D motions require high-accuracy astrometry and radial velocity information. Due to the large uncertainties caused by their dusty envelopes, their large angular sizes, and their surface brightness variability, AGB stars are problematic  targets for astrometry at optical wavelengths \citep[e.g.,][]{2014ARA&A..52..339R,2022ApJ...931...74S,2022A&A...667A..74A}. Furthermore, optical spectroscopic measurements can be significantly affected by the extinction at large distances. Such difficulties can be overcome by observing circumstellar masers at radio wavelengths, because maser velocities can provide accurate stellar radial velocities \citep[e.g.,][]{1995PASJ...47..815J}, and Very Long Baseline Interferometry (VLBI) astrometry of masers allows to measure the distances and proper motions of the maser-host stars with very high accuracy \citep[e.g.,][]{2012ApJ...744...23Z,2014ARA&A..52..339R}. 

Depending on the chemical properties of the stellar atmospheres, AGB stars can be divided into two main types: oxygen-rich (with O to C abundance ratio, [C/O], $<$1) and carbon-rich ([C/O]$>$1) \citep[e.g.,][]{2018A&ARv..26....1H}. While HCN and SiS masers are only detected toward a small number of C-rich stars \citep[e.g.,][]{2006ApJ...646L.127F,2017ApJ...843...54G,2018A&A...613A..49M,2022A&A...666A..69J}, SiO, H$_2$O, and OH masers have been found in the circumstellar envelopes of many thousands  O-rich stars \citep[e.g.,][]{1988ApJS...66..183E, 1991A&AS...90..327T, 1992A&AS...92...43L, 1992A&A...254..133L, 1996A&AS..116..117E, 1997A&AS..122...79S, 1998A&AS..128...35S, 2004PASJ...56..765D,2012JKAS...45..139K,2018MNRAS.473.3325W}, with SiO masers being the most detected. Their high maser detection frequency makes O-rich stars a potential tracer of the large-scale kinematics of the Milky Way. As a first step, one has to find masers in Galactic off-plane O-rich AGB stars for follow-up VLBI observations. 

It has long been known that the centroid of the velocity range covered by OH or SiO maser emission is a good measure of the host star's 
radial velocity \citep[e.g.,][]{Reid1976,1995PASJ...47..815J}. Because of better spectral resolution, radio or millimeter wavelength SiO observations usually allow higher accuracy radial velocity determinations than optical spectroscopy, for which radial velocity measurements depend on the variability phase of the star \citep{Scholz2000}. SiO masers are more frequently detected compared to H$_2$O masers in Mira-like AGB stars \citep[e.g.,][]{2014AJ....147...22K}. By far the most prominent OH maser transition found in evolved stars, the 1612 MHz line, because of its two horn profiles, is the best maser radial velocity tracer. However, it is mostly detected in higher mass-loss rate OH/IR stars, which are fewer in number than Mira-like O-rich AGB stars \citep{Habing1996}. Thus, SiO masers are ideal maser tracers of distant O-rich AGB stars.

SiO maser surveys have been extensively carried out in the Milky Way. A large number of AGB stars in the Galactic plane have been searched for SiO masers  in the 43 GHz SiO $v = 1 ~{\rm and}~2, J = 1-0$ transitions with the Nobeyama-45 m telescope \citep[e.g.,][]{1995PASJ...47..815J,2004PASJ...56..765D,2007ApJ...664.1130D,2007PASJ...59..559D,2010PASJ...62..525D}. Surveys for the 86 GHz SiO $v=1, $J$ = 2--1$ transition were  performed toward AGB stars in the inner Galaxy with the IRAM-30~m telescope \citep[e.g.,][]{2002A&A...393..115M,2018A&A...619A..35M}. These efforts have already led to the detection of roughly 2000 SiO masers in the Galactic plane. In recent years, the ongoing Bulge Asymmetries and Dynamical Evolution (BAaDE) survey\footnote{\url{https://leo.phys.unm.edu/~baade/index.html}} investigates $\sim$28000 SiO maser stars along the full Galactic plane, with the largest concentration in the Galactic bulge and inner Galaxy \citep[e.g.,][]{2018ApJ...861...75T,2018ApJ...862..153S,2019ApJS..244...25S,2020ApJ...892...52L}, which will dramatically increase the number of SiO masers in the Galactic plane. However, Galactic off-plane SiO masers which may originate in the thick disk, halo, and tidal stellar streams are much less explored.
Previous SiO and H$_2$O maser surveys of  Galactic off-plane AGB stars have led to the  detection of masers in the Galactic thick disk \citep[for example, ][]{2018MNRAS.473.3325W, 2022MNRAS.516.1881W}, and even a tentative detection of SiO maser emission in the Sagittarius tidal stream \citep{2007PASJ...59..559D}. However, the number ($\sim$40) of known masers could be located in the Galactic thick disk, which deviates from the Galactic plane by more than 1.2 kpc \citep{2018MNRAS.473.3325W}, is still too small to obtain reliable and statistically significant results on the kinematics and dynamics of this part of the Milky Way. Furthermore, a kinematical analysis of the 3D motions of masers found in the vicinity of the Sgr stellar streams suggests that the maser stars are not part of the stream \citep{2022MNRAS.516.1881W}, questioning the existence of masers in the tidal streams. Since these surveys are limited by their sensitivities, a deeper survey toward a well-selected sample has a higher possibility of detecting masers in distant Galactic structures. Therefore, we carried out a sensitive SiO maser survey toward a recently compiled sample of off-plane O-rich AGB stars that are believed to reside in the thick disk, the Galactic halo, and even the Sgr stream \citep{Mauron2019}. This study aims to shed light on the origin of these stars and their 3D motions, knowledge of which will contribute to our understanding of the formation history of the Milky Way and pave the way for follow-up observations of detected masers with VLBI, which will allow the determination of their trigonometric parallaxes and 3D motions. 

In this work, we report our discovery toward the first sample of 102 AGB stars. The sample and the observations are presented in Sect.~\ref{Sec:obs}. In Sect.~\ref{Sec:result}, we report the results of this survey. In Sect.~\ref{Sec:discuss}, we discuss the kinematics and origins of the SiO maser host stars. A summary is given in Sect.~\ref{Sec:sum}.


\section{Sample and Observations}\label{Sec:obs}

\subsection{Sample}

Recently, \cite{Mauron2019} compiled a catalog of 417 O-rich AGBs that are believed to reside in the thick disk, the Galactic halo, and the Sgr stellar stream.
These O-rich AGBs were collected from the variable catalogs of the LINEAR and Catalina optical monitoring surveys \citep{2013AJ....146..101P,2014ApJS..213....9D}, and over 95\% of them have the following properties: (i) 15$^\circ$ $< \left| b \right| <$ 90$^\circ$, where $b$ is Galactic latitude; (ii) Pulsation period between 90 and 400 days; (iii) ($K_s$)$_0 >$ 5.8 and 0.8 $<$ ($J-K_s$)$_0 <$ 2, that from 2MASS corrected for interstellar extinction; and (iv) Distance to the Sun between 4 and 80 kpc. 

For a pilot survey, our source selection strategies are: (1) a subset of 292 sources are initially considered based on their large distances ($>$5 kpc); We note that the distances from \cite{Mauron2019} are obtained through the $K_s$-band period-luminosity relation, and are likely overestimated (for our detections, see more details in Sect.~\ref{Sec:distance}).
(2) based on their Wide-field Infrared Survey Explorer (WISE) W4 band magnitudes, 292 sources were divided into two distinct groups, a bright group (2.5 $<$ W4 $<$ 7.5 mag) and a faint group (W4 $>$ 7.5 mag); 
(3) the final 102 sources are all selected from the  bright group, mainly based on right ascension.
Among them, 52 stars with right ascensions of $21^{\rm h}$--$5^{\rm h}$ were selected for observations with the Effelsberg-100 m observations, and 50 stars with right ascensions of $5^{\rm h}$--$12^{\rm h}$ were selected for the Tianma-65 m observations. 

\subsection{Characterization of our sample}\label{sec:Charact_sample}

\begin{figure*}[htbp]
\centering
\includegraphics[width=0.48\textwidth]{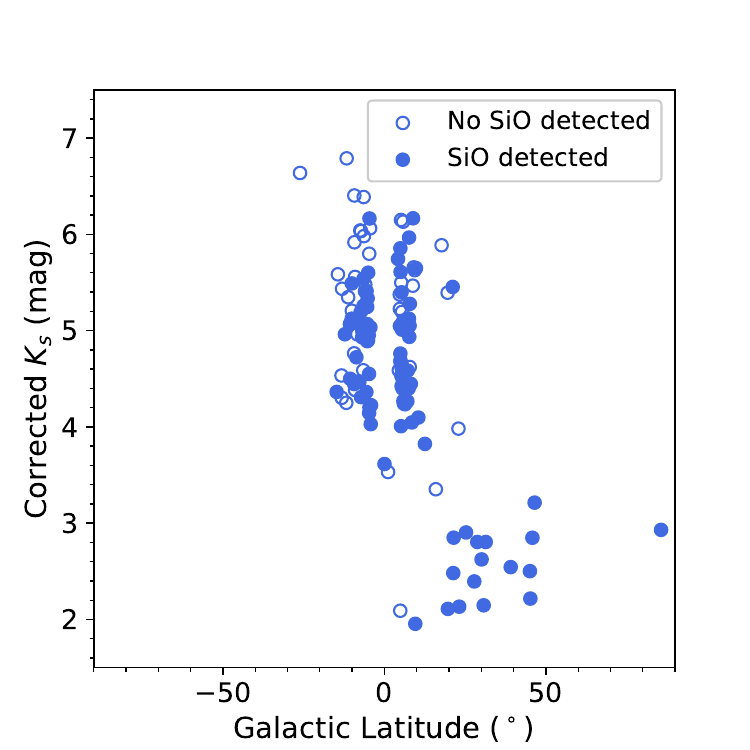}
\includegraphics[width=0.48\textwidth]{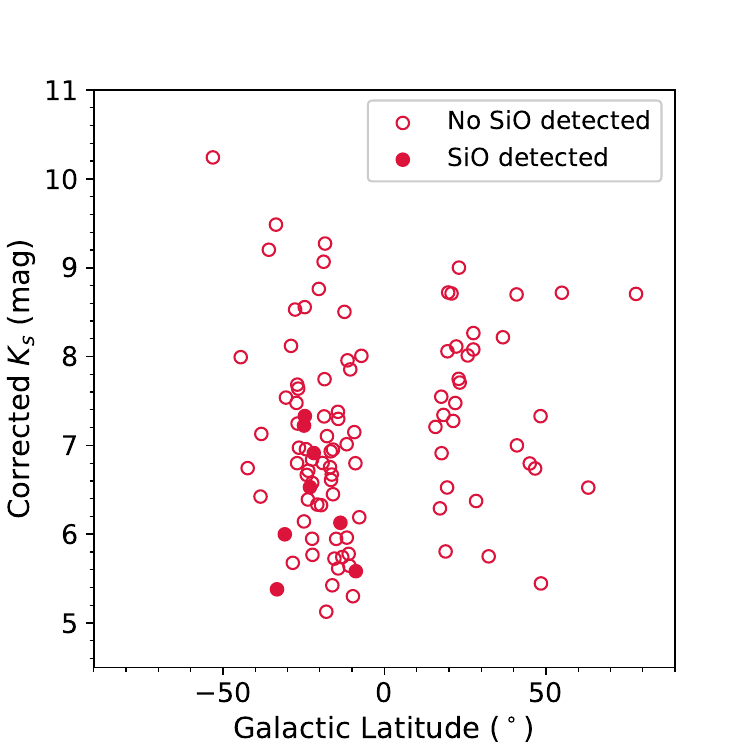}
\caption{Corrected 2MASS $K_s$ magnitude as a function of the Galactic latitude for the samples from \cite{2010PASJ...62..525D}(left) and our survey (right), respectively.
Filled and open circles indicate SiO maser detections and non-detections, respectively.
In the right panel, a data point with corrected $K_s$ of 13.8 mag and Galactic latitude of 34$^\circ$ (with no SiO maser detected) is not shown, in order to better visualize the distribution of the sample.
\label{fig:ks}}
\end{figure*}

We characterized our sample through a comparative analysis with another off-plane SiO maser survey conducted by \cite{2010PASJ...62..525D}.
Fig.~\ref{fig:ks} shows the corrected 2MASS $K_s$ magnitude as a function of the Galactic latitude for that survey's sample 
and our sample, respectively. 
In Galactic latitude distribution, our AGB star targets cover a broader latitude range (7$^\circ$ $< | b | <$ 80$^\circ$) than the sample of \citet{2010PASJ...62..525D}.
In addition, our 102 targets are generally fainter with corrected $K_s$ magnitudes between 5 and 10.5 mag, while those observed by \citet{2010PASJ...62..525D} range from 2 to 7 mag.
At mid-infrared wavelengths, our sources are also fainter.
All targets from \cite{2010PASJ...62..525D} have Infrared Astronomical Satellite (IRAS) counterparts in the IRAS Point Source Catalog (PSC) with 12~$\mu m$ fluxes ($F_{\rm 12}$) in a range of 2.3 -- 63~Jy (with a median of 7.5~Jy). While only 21 stars out of our 102 targets are seen in the IRAS PSC with flux densities of $0.27$\,Jy $ < F_{\rm 12} < 15$\,Jy (with a median of 0.6~Jy), which are much lower than those of the \citet{2010PASJ...62..525D} sample. 

\begin{figure}[htbp]
\centering
\includegraphics[width=0.5\textwidth]{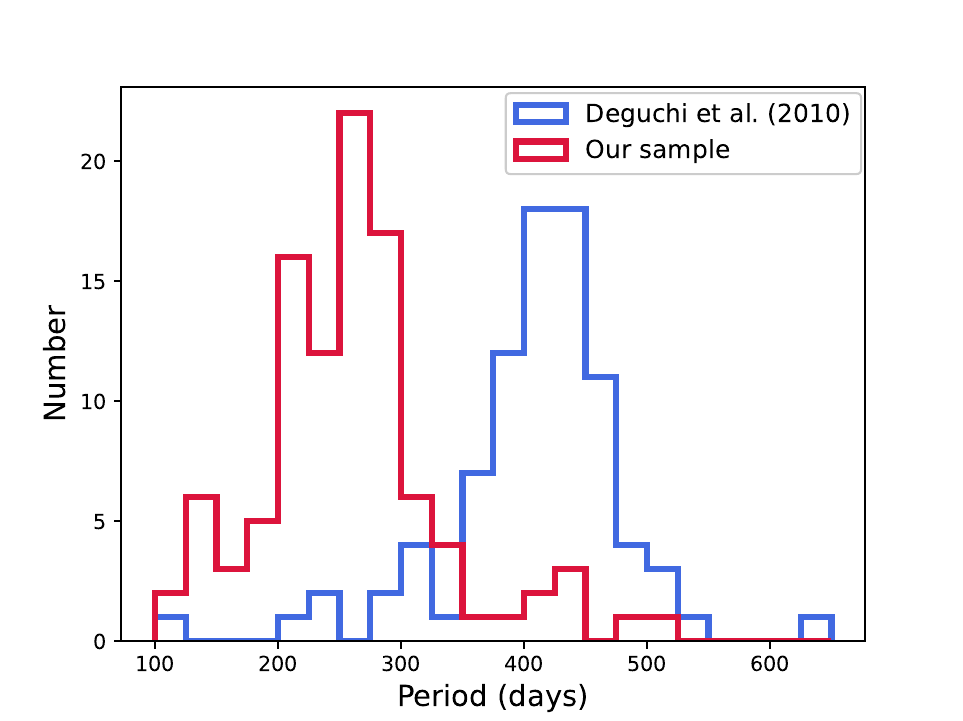}
\caption{Period distribution of our sample (red) and the sample of \citealt{2010PASJ...62..525D} (blue).
\label{fig:period}}
\end{figure}

Figure~\ref{fig:period} shows the distribution of pulsation periods for the two samples, and it reveals that the periods of the stars in our sample are typically shorter than those of the \citet{2010PASJ...62..525D} sample.
The periods of the stars observed by \citet{2010PASJ...62..525D} were derived from Zwicky Transient Facility (ZTF) light curves\footnote{\url{https://irsa.ipac.caltech.edu/Missions/ztf.html}}. 
The median values of periods for our sample and the \citet{2010PASJ...62..525D} sample are 260 and 415 days, respectively.
In addition, we performed a Kolmogorov-Smirnov test, which suggests that the periods of the two samples are significantly different indicated by a very small $p$-value ($\ll$0.0013).
Given that period is a relatively good indicator of age \citep[see details in ][]{2023arXiv230617758T}, we infer that our targets are generally older than the targets in \cite{2010PASJ...62..525D}.

\begin{figure*}[htbp]
\centering
\includegraphics[width=0.95\textwidth]{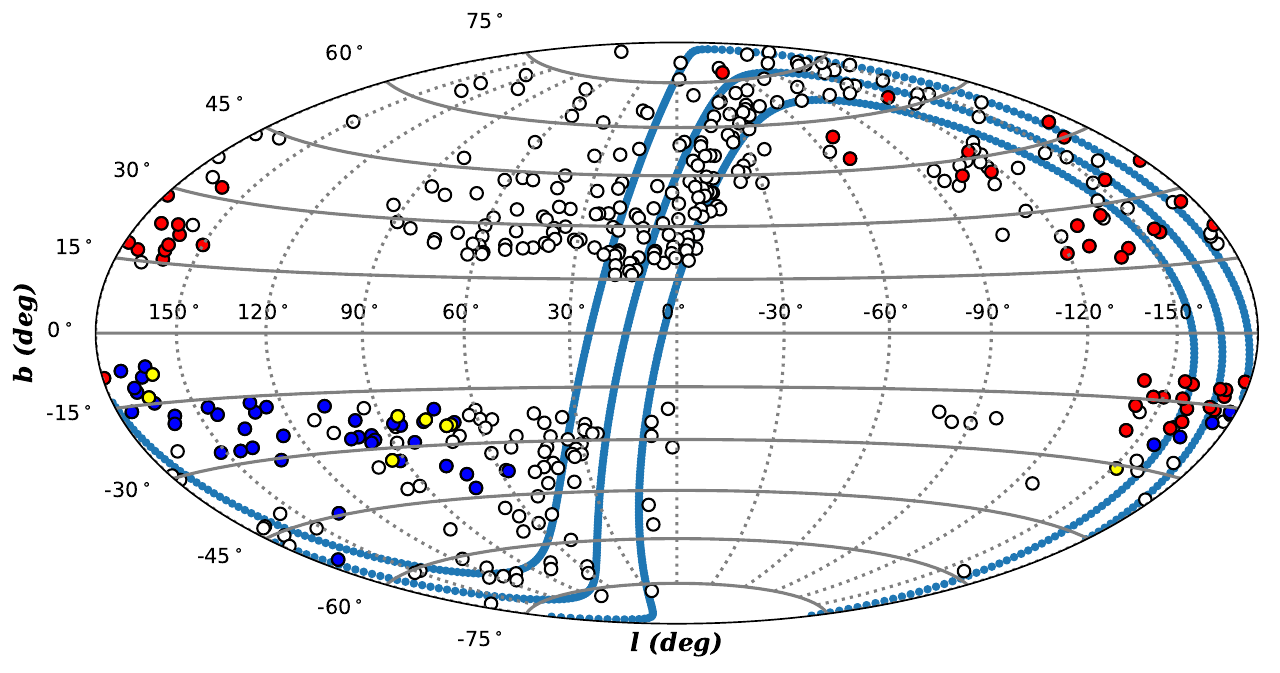}
\caption{
Galactic locations of the 417  O-rich AGB stars from \cite{Mauron2019}. The three blue lines indicate the $B$ = 0$^\circ$, $\pm$10$^\circ$ planes, where $B$ is the latitude of Sgr stream coordinate system. The red circles denote 50 sources observed with the Tianma-65 m telescope, the blue circles denote 52 sources observed with the Effelsberg-100 m telescope. Yellow circles denote the 8 sources with the detection of SiO masers.  The open circles represent the stars which are not included in the current pilot survey.
\label{fig:260stars}}
\end{figure*}

\subsection{Effelsberg-100~m Telescope Observations}
We conducted Q-band observations of the selected 52 stars using the Effelsberg-100 m telescope\footnote{\url{https://www.mpifr-bonn.mpg.de/effelsberg/astronomers}} \citep{1973IEEEP..61.1288H,2011JAHH...14....3W} to observe SiO $J =1-0, v = 1~\rm{and}~2$ transitions (rest frequencies at 43.122030 and 42.820480~GHz, \citealt{2005JMoSt.742..215M}) in 2022 September and December, 2023 January and February (project ID: 20-22). The observations were performed in position switching mode with the off-position at an offset of 10$\arcmin$ in right ascension. 
The dual-polarization S7mm Double Beam RX secondary focus receiver\footnote{\url{https://eff100mwiki.mpifr-bonn.mpg.de/doku.php?id=information_for_astronomers:rx_list}} was used as the front end.
Fast Fourier Transform Spectrometers (FFTSs) were used as the backend to record signals, and each FFTS has an instantaneous broad bandwidth of 2.5 GHz with 65536 channels, yielding a channel width of 38.1 kHz and a corresponding velocity spacing of 0.27~\kms\,(multiply by 1.16 to convert to velocity resolution, see \citealt{2012A&A...542L...3K}). This setup affords a broad frequency coverage of 39.5--44.5~GHz, which allows us to observe the SiO $J =1-0, v =0$--3 transitions simultaneously.

NGC 7027 was used for pointing, focus, and flux density calibration at the start of each observation session.
We verified the pointing every hour on nearby strong continuum or SiO maser sources. The pointing accuracy was about 5$\arcsec$ at 43~GHz. The half-power beam width (HPBW) is about 20$\arcsec$ with a main beam efficiency of 39\% at 43~GHz. The antenna efficiency is 24\%, corresponding to a factor of 1.47 Jy~K$^{-1}$ to convert antenna temperature into flux density.
The system temperature ranged from 160 to 280~K during the observations.
An on-source integration time of 10~minutes was used for the initial search of SiO masers toward each source, which achieves a 1$\sigma$ noise level of approximately 40--60~mJy at a channel width of 0.27~\kms.  
Supplementary observations were carried out for the sources with possible SiO maser detection to improve the signal-to-noise (S/N) ratio, which reduces the 1$\sigma$ noise level to about 20--30~mJy at a channel width of 0.27~\kms.
The Effelsberg observations took about 38 hours in total.

\subsection{Tianma-65 m Observations}

We conducted Q-band observations of 50 stars using the Tianma-65 m telescope \citep{2015ApJ...814....5Y} to observe the SiO $J=1-0,\, v =1~{\rm and}~2$ transitions in 2022 December and 2023 January.
The observations were performed in position switching mode with the off-position at an offset of 30$\arcmin$ in azimuth.
The dual-polarization and dual-beam Q-band cryogenic receiver \citep{2018RAA....18...44Z} and the Digital Backend System \citep[DIBAS;][]{2012AAS...21944610B} were used to receive and record signals, with DIBAS being a field-programmable gate array (FPGA)-based spectrometer. For spectral line observations, DIBAS supports 29 observing modes with different frequency bandwidths and resolutions. In our observations, only beam 2 was employed to track targets and the spectrometer worked in Mode-3, which provides two frequency banks for both polarization, centered at 42.830~GHz and 43.132~GHz, respectively. Each frequency bank has a bandwidth of 500~MHz with 16384 channels, yielding a channel spacing of 30.5 kHz and a corresponding velocity spacing of 0.21~km~s$^{-1}$.


During the observations, the receiver noise temperatures were roughly 30--40~K, and the system temperature ranged from 85 to 135 K.
Pointing was regularly checked 
to ensure the accuracy of better than 10$\arcsec$. The HPBW at 43~GHz is $\sim$30$\arcsec$.
The main beam efficiency is $\sim$60\%\,depending on the elevation.
Since the telescope has an active surface control system to correct gravity-caused deformation of the main reflector during observations \citep{2018ITAP...66.2044D}, the Q-band aperture efficiency remains identical (0.5$\pm$0.1) at elevations in a range of 15$^\circ$--80$^\circ$ \citep{2018RAA....18...44Z}. Thus, the corresponding factor of 1.67~Jy~K$^{-1}$ is adopted to convert antenna temperature to flux density \citep{2017AcASn..58...37W}. 
An on-source integration time of 20 minutes was used for all 50 sources, which achieves 1$\sigma$ noise level of approximately 20 -- 40 mJy, which are very close to those obtained with the Effelsberg-100~m telescope. The Tianma-65 m observations took about 40 hours in total. 


\subsection{Data reduction}
The data were processed using the GILDAS/CLASS package \citep{2005sf2a.conf..721P}.
Although both the Effelsberg-100~m and Tianma-65 m telescopes have two beams, we only analyzed the data for the one that tracked target positions. A low-order ($<$3) polynomial baseline subtraction was performed for each spectrum. For sources observed using the Effelsberg-100~m in two epochs, we averaged their spectra to achieve higher S/N ratios. Velocities are given with respect to the local standard of rest (LSR) throughout this work.

\section{Results} \label{Sec:result}
Out of the 102 targets, toward eight sources we detected SiO maser emission in at least one transition 
above 3$\sigma$ level. Thus, the detection rate of our SiO maser survey was 8$\pm$3\%, where the uncertainty is estimated by assuming a binomial distribution. 
The detection rate of our survey is lower than those of previous off-plane SiO maser surveys, which were 63\%\ for that of \cite{2010PASJ...62..525D} and 20\%\ for that of \cite{2018MNRAS.473.3325W}. The reason for this could be that our targets are fainter, and harbor shorter pulsation periods (i.e., older; see details in Sect.~\ref{sec:Charact_sample}).
Fig.~\ref{fig:ks} also suggests that SiO maser emission can hardly be detected in faint stars with corrected $K_s$ magnitude $>$ 7.5 mag at our current sensitivity.

After cross-matching the Galactic SiO maser catalog from \cite{2018MNRAS.473.3325W}, we found that all of our eight SiO masers were detected for the first time. In addition, all these masers were detected using the Effelsberg-100 m telescope. The source information and 1$\sigma$ noise level of the 94 non-detections in our survey are listed in Table~\ref{Tab:non-det}. In addition, the SiO $J = 1-0, v =0~{\rm and}~ 3$ lines were observed but not detected with the Effelsberg-100 m telescope, with  3$\sigma$ upper limits of 60--180~mJy. 

Figure~\ref{fig:260stars} shows the distribution of the stars with detected SiO masers (marked with yellow circles) and observed stars in the plane of the sky.
The red circles denote 50 non-detections observed with the Tianma-65 m, the blue circles denote 44 non-detections observed using the Effelsberg-100 m.
The open circles represent the stars from \citet{Mauron2019}, which are not included in the current pilot survey.

\begin{figure*}[!htbp]
\centering
\includegraphics[width=0.95\textwidth]{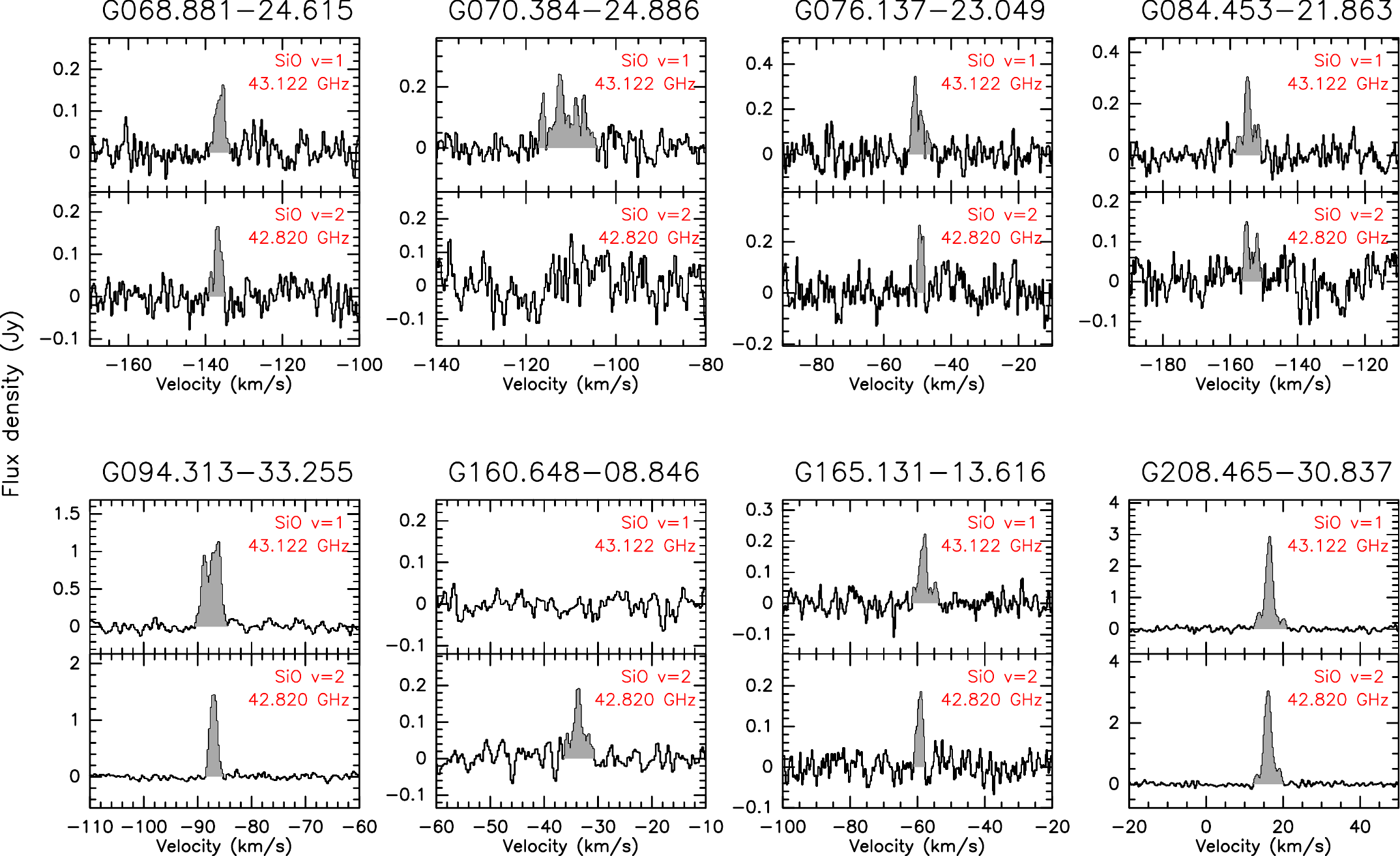} 
\caption{SiO maser spectra toward the eight stars with detections. 
The gray-shaded regions indicate the velocity range used to determine the integrated intensity. 
Spectra observed in multiple epochs were averaged in order to improve the S/N ratios.
\label{fig:det}}
\end{figure*}

The SiO maser spectra of the eight sources are shown in Fig.~\ref{fig:det}.
All detections show distinct singular or multiple narrow features, and the line profiles are completely different from the broad, smooth profiles of thermally excited emission lines commonly detected in evolved stars \citep[e.g.,][]{1995ApJ...445..872Y,2022A&A...666A..69J}.
Thus, all of the detected signals can be regarded as SiO maser emission, especially for the sources with velocity-aligned emission in both transitions.
Based on the spectra, we derive the lines' peak LSR velocities, velocity ranges, integrated intensities, and peak intensities, which are listed in Table~\ref{Tab:det}.

The APOGEE DR17 \citep{2022ApJS..259...35A}, RAVE DR6 \citep{2020AJ....160...82S} and Gaia DR3 \citep{2023gaiadr3} catalogs do not provide the radial velocities of these SiO maser-host O-rich stars. Given that SiO masers can be used as a good indicator of stellar radial velocity \citep[e.g.,][]{1995PASJ...47..815J}, we determine the radial velocities of these stars in the LSR frame (i.e., $V_{\rm LSR}$ in Table~\ref{Tab:kin}) for the first time, utilizing the peak velocity of SiO masers as a proxy.
For the sources detected in both SiO maser transitions, we take the average of these maser peak velocities as the stellar radial velocity for the analyses in Sect.~\ref{sec:kin}. 


Toward six sources, both SiO $J =1-0, v =1~{\rm and}~2$ masers are detected, while we only detect the SiO $J =1-0, v =1$ maser line toward G070.384$-$24.886 and only the SiO $J =1-0, v =2$ maser line toward G160.648$-$08.846. 
Generally, the $v$=1 masers are slightly brighter than their $v$=2 counterparts, except for G160.648$-$08.846 and G208.465$-$30.837.
For the six sources that harbor both $v$=1 and 2 masers, the velocities of the $v$=1 and 2 masers are in good agreement, which is consistent with previous SiO maser surveys \citep[e.g.,][]{1995PASJ...47..815J,2018MNRAS.473.3325W}.
The integrated intensity ratios between the $v = 1$ and $v = 2$ masers range from 0.94 to 2.05, which are comparable to the expected values of the SiO--H$_{2}$O overlap pumping model \citep[see Fig.~4 in][]{2014A&A...565A.127D}.

Previous observations and theoretical studies suggest that the SiO masers in Mira variables reach their maximum flux densities at an optical phase of 0.1--0.25, which corresponds to the infrared maximum  \citep[e.g.,][]{2004A&A...424..145P,2009MNRAS.394...51G}. 
Most epochs of SiO maser emission we detected correspond to the optical phases of 0.08–0.33 (see Column 5 in Table~\ref{Tab:det}), which is in line with a radiative pumping scenario. This indicates that observations around the optical phase of 0.1--0.25 can increase the probability of detecting new SiO masers toward O-rich stars, in particular for distant targets.

We estimate the isotropic luminosity, or ``photon rate'', of SiO masers using the following equation: 
\begin{equation}
L_{\rm SiO} = \frac{4\pi D_{\rm adopt}^2 \int S{\rm d}v}{hc},
\end{equation}
where $D_{\rm adopt}$ is the distance that we adopted in this work (see details in Sect.~\ref{Sec:distance}), $\int S{\rm d}v$ is the velocity-integrated flux density of SiO maser emission, $h$ is Planck’s constant, and $c$ is the speed of light. 
In convenient units this translates to:
\begin{equation}
L_{\rm SiO} ({\rm photons\,s}^{-1}) = 6.04\times 10^{41}~D_{\rm adopt}^2({\rm kpc}) \int S{\rm d}v ({\rm Jy\,km~s}^{-1}),
\end{equation}
Column 13 of Table~\ref{Tab:det} lists the SiO maser luminosity of each transition for individual sources.
In comparison to SiO masers generally found in evolved stars, the maser luminosities of our detections are comparable or slightly lower than those of Miras (a few $\times$~10$^{43}$ photons s$^{-1}$, \citealt{2014AJ....147...22K}).

\begin{deluxetable*}{lcccccccccccc} 
\tablecaption{Observational results for the detected sources}. \label{Tab:det}
\centering
\tabletypesize{\small}
\tablewidth{1pt}
\tablecolumns{13}
\tablenum{1}
\renewcommand\arraystretch{1.1}
\centering
\setlength{\tabcolsep}{2pt}
\tablehead{
\colhead{Name} & \colhead{$\alpha_{\rm J2000}$} & \colhead{$\delta_{\rm J2000}$} & \colhead{epoch} & \colhead{$\phi$} & \colhead{Period} & \colhead{Transition} & \colhead{$V_{\rm pk}$} & \colhead{$V_{\rm range}$} & \colhead{$\int S{\rm d}v$} & \colhead{$S_{\rm pk}$} & \colhead{1$\sigma$} & \colhead{$L_{\rm SiO}$} \\
 &  &  &  &  & \colhead{(d)} &  & \colhead{(\kms)} & \colhead{(\kms)} & \colhead{(Jy~\kms)} & \colhead{(Jy)} & \colhead{(Jy)} & \colhead{(photons s$^{-1}$)} } 
\startdata
G068.881$-$24.615 & 21:30:47.88 & +16:33:49.1   & B, F & 0.30, 0.80 & 307.7 & $v$=1 & $-$135.4 & [$-$138.7,$-$133.4] & 0.41 & 0.16 & 0.02 & 6.4$\times$10$^{42}$\\
                  &             &               &      &            &       & $v$=2 & $-$136.6 & [$-$139.0,$-$135.1] & 0.34 & 0.16 & 0.03 & 5.3$\times$10$^{42}$\\
\hline
G070.384$-$24.886 & 21:35:26.69 & +17:24:06.7   & B    & 0.26       & 336.7 & $v$=1 & $-$112.6 & [$-$117.0,$-$104.5] & 1.24 & 0.24 & 0.03 & 2.3$\times$10$^{43}$ \\
\hline
G076.137$-$23.049 & 21:45:17.86 & +22:27:56.0   & A    & 0.21       & 251.2 & $v$=1 & $-$50.6  & [$-$52.4,$-$46.3]   & 0.86 & 0.34 & 0.05 & 1.0$\times$10$^{43}$\\
                  &             &               &      &            &       & $v$=2 & $-$49.4  & [$-$50.0,$-$48.1]   & 0.42 & 0.26 & 0.05 & 5.1$\times$10$^{42}$\\
\hline
G084.453$-$21.863 & 22:07:09.89 & +28:28:37.4   & A    & 0.32       & 264.5 & $v$=1 & $-$154.8 & [$-$158.2,$-$150.9] & 0.86 & 0.30 & 0.03 & 1.3$\times$10$^{43}$\\
                  &             &               &      &            &       & $v$=2 & $-$155.1 & [$-$156.4,$-$150.7] & 0.43 & 0.15 & 0.04 & 6.7$\times$10$^{42}$\\
\hline
G094.313$-$33.255 & 23:07:36.31 & +23:46:56.6   & C    & 0.33       & 280.8 & $v$=1 & $-$86.1  & [$-$90.3,$-$84.6]   & 3.66 & 1.12 & 0.05 & 1.3$\times$10$^{43}$\\
                  &             &               &      &            &       & $v$=2 & $-$87.0  & [$-$88.2,$-$85.6]   & 2.37 & 1.45 & 0.04 & 8.2$\times$10$^{42}$\\
\hline
G160.648$-$08.846 & 04:14:34.50 & +38:43:16.7   & D, E & 0.12, 0.21 & 275.4 & $v$=2 & $-$33.5  & [$-$36.3,$-$31.0]   & 0.42 & 0.19 & 0.02 & 2.1$\times$10$^{42}$ \\
\hline
G165.131$-$13.616 & 04:13:30.34 & +32:14:18.4   & D, E & 0.08, 0.16 & 287.1 & $v$=1 & $-$57.7  & [$-$61.4,$-$53.9]   & 0.64 & 0.22 & 0.03 & 1.3$\times$10$^{43}$\\
                  &             &               &      &            &       & $v$=2 & $-$58.9  & [$-$61.0,$-$57.9]   & 0.36 & 0.19 & 0.02 & 7.0$\times$10$^{42}$\\
\hline
G208.465$-$30.837 & 04:52:57.25 & $-$10:01:58.8 & D    & 0.08       & 425.2 & $v$=1 & $+$16.5  & [$+$12.7,$+$20.6]   & 7.11 & 2.93 & 0.06 & 3.4$\times$10$^{43}$\\
                  &             &               &      &            &       & $v$=2 & $+$16.1  & [$+$12.5,$+$20.0]   & 7.59 & 3.04 & 0.05 & 3.6$\times$10$^{43}$\\
\enddata
\tablecomments{Columns 1--3 give the source name and coordinates. Column 4 gives the  observing dates (yyyy-mm-dd), A: 2022-09-02, B: 2022-09-08, C: 2022-09-13, D: 2022-12-18, E: 2023-01-11, F: 2023-02-07. The corresponding optical phases and periods derived from the ZTF light curves are presented in Columns~5--6. Columns 7--12 list the detected SiO transition, the peak velocity $V_{\rm pk}$, velocity range $V_{\rm range}$, integrated flux density $\int S {\rm d}v$, peak intensity $S_{\rm pk}$ and 1$\sigma$ noise level, respectively. 
Column 13 lists the isotropic photon luminosity of SiO masers. 
}
\end{deluxetable*}

\section{Discussion}\label{Sec:discuss}

\subsection{Distance and Galactic distribution} \label{Sec:distance}

Distance is a key parameter in studying the Galactic location and kinematics of our targets. The catalog of O-rich AGBs targeted in our study was compiled by \citet{Mauron2019}, who adopted the $K_s$-band period-luminosity relationship (PLR) to determine the distances. However, the distance accuracy is limited by the uncertainties of the 2MASS $K_s$ magnitude, periods, and the scatter of PLR. 


\cite{2018MNRAS.473.3325W} utilized two methods (i.e., $K_s$-band PLR and WISE luminosity, see details in their Appendix A) to estimate the distances of O-rich AGB stars in their sample. They find that distances estimated by these two methods agree within 2~kpc. However, due to the potentially more severe extinction in near-IR bands, the distances estimated by the $K_s$-band PLR are much larger than those derived from the WISE luminosity for distant stars. \cite{2022MNRAS.516.1881W} suggested that the uncertainties of the $K_s$-band PLR distances could be larger than 30\%\, or even higher.

Here, we revisit the distances of our target stars, employing two different methods.
Firstly, we investigate the distances of these sources from the Gaia DR3 \citep[][]{2023gaiadr3}. Due to the bias and underestimation of parallax uncertainties in the Gaia DR3 catalog \citep[e.g.,][]{2021A+A...649A...4L,2021A&A...649A...5F}, we make the zero point correction of parallaxes \citep{2021A+A...649A...4L}, and amend the parallax bias by applying the method of \citet{2022A+A...657A.130M}. 
Secondly, we derive the distances by adopting the WISE W1, W2, and W3 band PLRs ($D_{\rm m-PLR}$) \citep{2021ApJ...919...99I} which were calibrated with Miras in the Large Magellanic Cloud. 
The reliability of these mid-IR PLRs is verified using  Galactic O-rich Miras \citep{2023ApJS..264...20I}. The mid-IR PLRs are less affected by the extinction than the $K_s$-band PLR, making them a more reliable and independent tool to estimate the distances to distant stars.
Taking their uncertainties into account, we find that the mid-IR PLR distances are generally consistent with Gaia trigonometric parallax distances within the respective errors for the maser-host stars in our sample. 
Hence, we take an error-weighted average between Gaia trigonometric distance and mid-IR PLR distance to obtain the distance ($D_{\rm adopt}$) adopted for this study. We note that the adopted distances in our work are generally 60\% of the distances listed in the catalog of \cite{Mauron2019}, indicating that our stars lie at closer distances than previously thought.
Table~\ref{Tab:kin} lists the corrected Gaia DR3 parallax, the mid-IR PLR distance, the adopted distance, the Galactocentric distance, and the distance to the Galactic plane for every source.

\begin{figure*}[!htbp]
\centering
\includegraphics[width=0.65\textwidth]{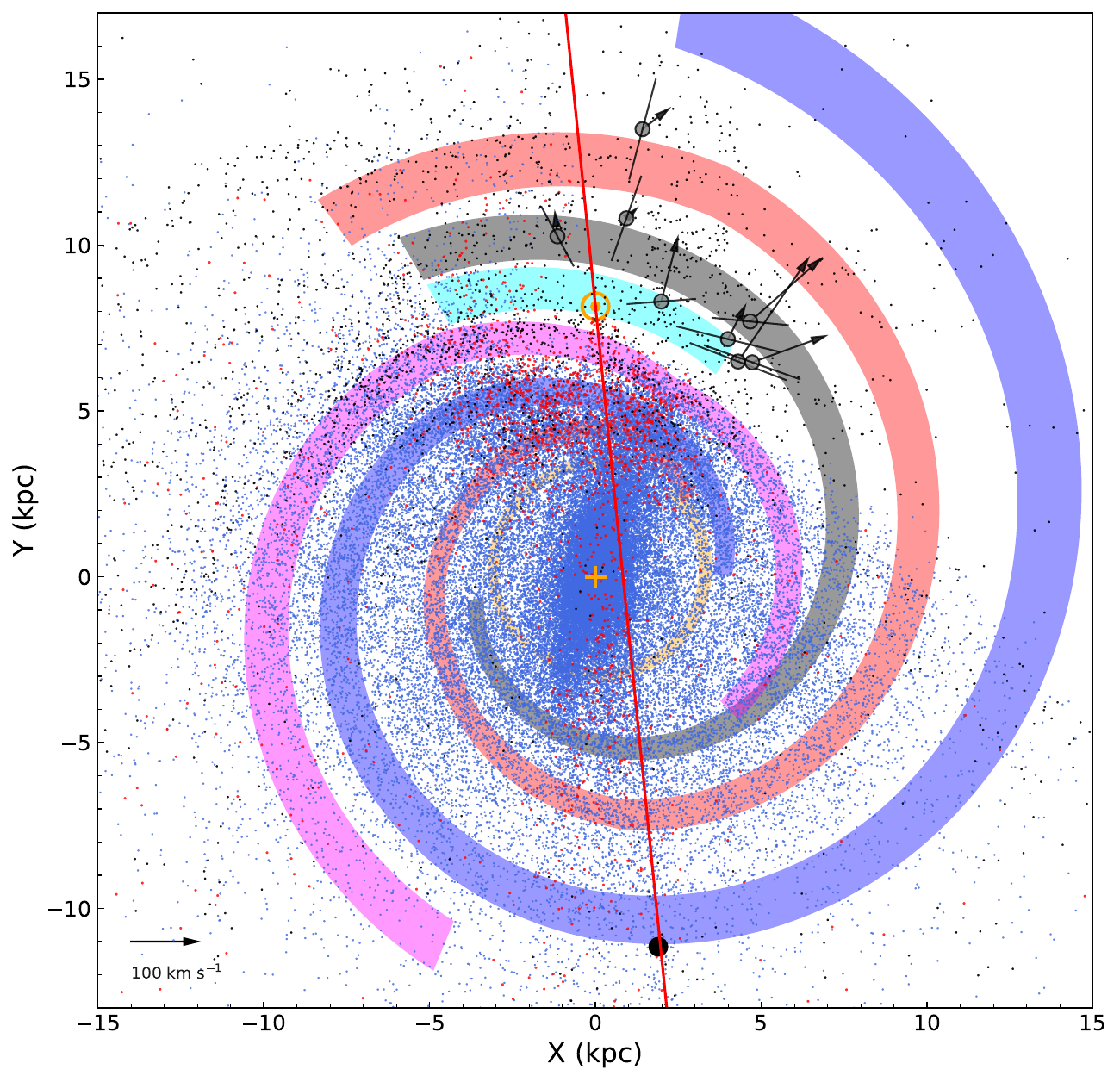}
\includegraphics[width=0.65\textwidth]{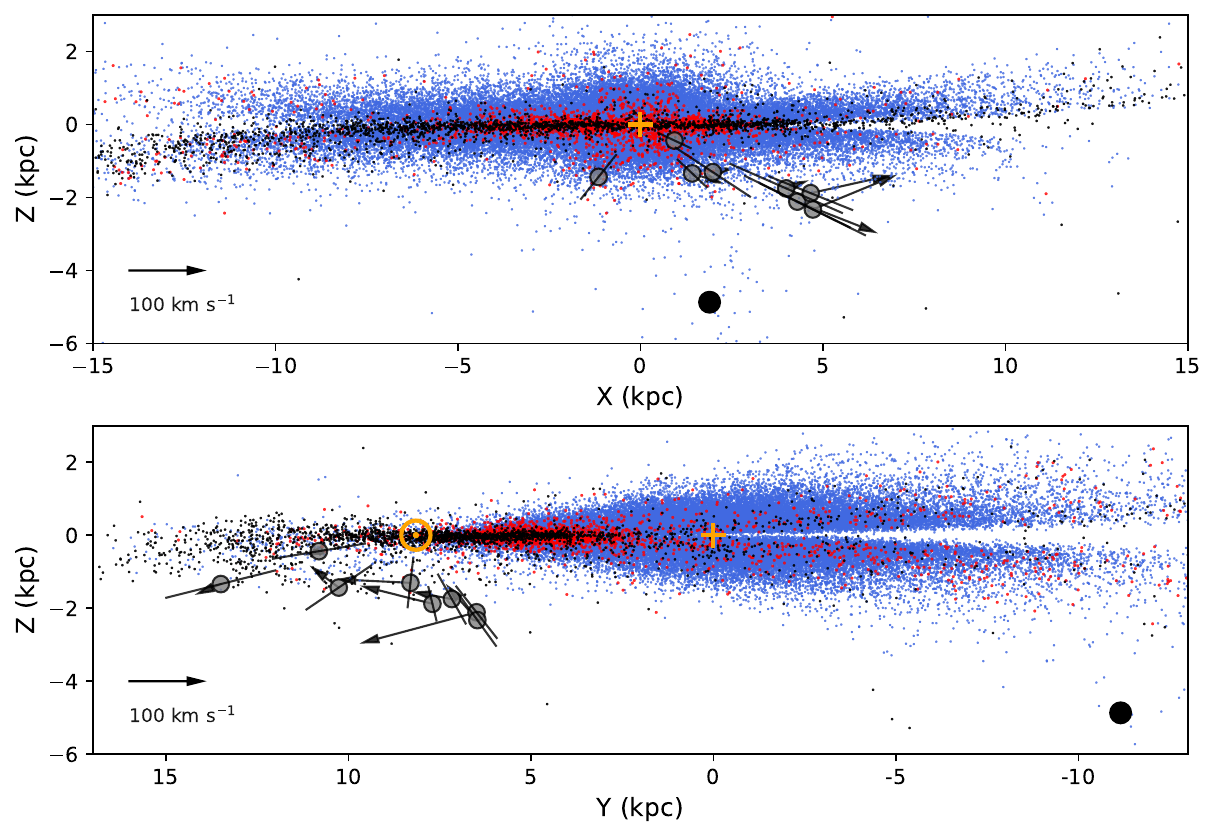}
\caption{Two-dimensional projections in the Galactocentric Cartesian coordinates, with the Sun at (X,Y,Z) = (0 kpc, 8.15 kpc, 0 kpc) (orange circle with dot). The Galactic center is marked as an orange plus. The gray circles denote the SiO masers found in the survey, whose arrows depict the projections of their peculiar motions.
The location of the Galactic plane is delineated using O-rich and C-rich Miras from \citet{2023ApJS..264...20I}, represented by blue and red dots, respectively, while classical Cepheids from \citet{2019Sci...365..478S} are indicated by black dots. The presence of the warped disk is evident in the middle panel (for more details, see \citealt{2019Sci...365..478S} and \citealt{2023ApJS..264...20I}). 
The filled black circle marks the positions of the Sagittarius dwarf spheroidal galaxy, which is located at a distance of $\sim$20~kpc \citep{2004AJ....127.2031K} in the direction of ($l$, $b$) = (5$^\circ$.608, $-$14$^\circ$.086). 
The X-Y plot incorporates the Galactic spiral arm model, based primarily on \citet{2019ApJ...885..131R}, as well as an updated structure of the Sagittarius arm that contains new maser parallax measurements (Bian et al. in prep).
The red line in the top panel indicates the intersections between the Sgr orbital plane and the Galactic mid-plane.
For a 3D interactive view,  \href{https://gongyan2444.github.io/3D/mw-sio.html}{click me}. 
\label{fig:3dplot}}
\end{figure*}

Using the distance information, we derive the Galactic distribution of our sources as shown in Fig.~\ref{fig:3dplot}. 
From the projection of the Galactic plane, as shown in the top panel of Fig.~\ref{fig:3dplot}, G165.131$-$13.616 is located far beyond the Norma–Outer arm, G084.453$-$21.863, G160.648$-$08.846 and G208.465$-$30.837 are in the Perseus arm, G070.384$-$24.886, G076.137$-$23.049 and G094.313$-$32.255 coincide in the Local arm, while G068.881$-$24.615 appears to be located near the inner side of the Perseus arm. 
However, the SiO masers are detected in off-plane stars which are unlikely associated with the Galactic spiral arms. This is expected, because low-mass evolved stars including these maser-host AGB stars are not reliable indicators of spiral arms \citep[][Bian et al. in prep; see also Fig.~\ref{fig:3dplot}]{2019ApJ...885..131R}.

From the side view of the Galaxy (see the middle and bottom panels in Fig.~\ref{fig:3dplot}), the stars, except for G160.648$-$08.846,  are significantly below the Galactic plane with $|Z|=$ 1.3--2.3~kpc.
Compared with the scale heights of the Galactic thin and thick disks which range from $\sim$120 to 300 pc, and $\sim$500 to 1400 pc, respectively \citep[e.g.,][]{1983MNRAS.202.1025G,2008ApJ...673..864J,2010ApJ...714..663D}, the small $|Z|$ of G160.648$-$08.846 indicates this source is probably located in the Galactic thin disk, while the large $|Z|$ of the others  indicate that they are not located in the thin disk of the Milky Way, but probably in the thick disk instead.


We also note that the adopted distance uncertainties can be as high as 50\% in our sample. Such parameters can be improved by obtaining accurate stellar distances by directly measuring the parallax of the SiO masers \citep[e.g.,][]{2012ApJ...744...23Z}.

\begin{deluxetable*}{lcccrcrrrrrr}
\tablecaption{Distance and kinematic information of the stars with SiO maser detection.\label{Tab:kin}}
\centering
\tabletypesize{\scriptsize}
\tablewidth{1pt}
\tablecolumns{12}
\tablenum{2}
\renewcommand\arraystretch{1.05}
\centering
\setlength{\tabcolsep}{1.6pt}
\tablehead{
\colhead{Name} & \colhead{Parallax} & \colhead{$D_{\rm m-PLR}$} & \colhead{$D_{\rm adopt}$} & \colhead{$R$}  & \colhead{$Z$} & \colhead{$\mu_{\rm x}$} & \colhead{$\mu_{\rm y}$} & \colhead{$V_{\rm LSR}$} & \colhead{$U_{\rm s}$} & \colhead{$V_{\rm s}$} & \colhead{$W_{\rm s}$}  \\
\colhead{} & \colhead{(mas)} & \colhead{(kpc)} & \colhead{(kpc)} & \colhead{(kpc)}  & \colhead{(kpc)} & \colhead{(mas yr$^{-1}$)} & \colhead{(mas yr$^{-1}$)} & \colhead{(\kms)} & \colhead{(\kms)} & \colhead{(\kms)} & \colhead{(\kms)}  } 
\startdata
G068.881$-$24.615 & 0.0792$\pm$0.1530 & 4.79$\pm$1.07  & 5.1$\pm$1.7  &  8.1$\pm$0.7 & $-$2.1$\pm$0.7 & 1.301$\pm$0.076    & $-$6.217$\pm$0.066 & $-$136.0 & $-$184.61$\pm$ 7.50 & $-$2.73$\pm$18.22 & $-$33.52$\pm$27.03 \\
G070.384$-$24.886 & 0.2647$\pm$0.1521 & 6.42$\pm$1.63  & 5.5$\pm$1.7  &  8.3$\pm$0.7 & $-$2.3$\pm$0.7 & $-$4.021$\pm$0.086 & $-$5.380$\pm$0.082 & $-$112.6 & $-$92.18$\pm$31.43  & 69.83$\pm$20.85   & 44.46$\pm$ 5.30   \\
G076.137$-$23.049 & 0.3277$\pm$0.3549 & 4.93$\pm$1.13  & 4.5$\pm$1.7  &  8.4$\pm$0.7 & $-$1.8$\pm$0.7 & $-$2.345$\pm$0.108 & $-$3.974$\pm$0.098 & $-$50.0  & $-$56.83$\pm$ 7.42  & $-$2.36$\pm$ 3.14 & 11.07$\pm$ 6.20   \\
G084.453$-$21.863 & 0.2581$\pm$0.0869 & 5.96$\pm$1.33  & 5.1$\pm$1.2  &  9.2$\pm$0.6 & $-$1.9$\pm$0.5 & $-$1.759$\pm$0.064 & $-$4.104$\pm$0.056 & $-$154.9 & $-$128.78$\pm$14.34 & 47.61$\pm$10.26   & 17.56$\pm$11.18  \\
G094.313$-$33.255 & 0.3735$\pm$0.2583 & 2.27$\pm$0.66  & 2.4$\pm$1.2  &  8.6$\pm$0.5 & $-$1.3$\pm$0.7 & 1.580$\pm$0.100    & $-$5.730$\pm$0.098 & $-$86.6  & $-$102.64$\pm$12.92 & 1.02$\pm$ 5.44    & $-$2.80$\pm$27.72  \\
G160.648$-$08.846 & 0.6016$\pm$0.1818 & 4.06$\pm$0.98  & 2.9$\pm$1.4  & 10.9$\pm$1.3 & $-$0.4$\pm$0.2 & 0.185$\pm$0.150    & $-$1.859$\pm$0.091 & $-$33.5  & $-$15.14$\pm$ 7.35  & 13.79$\pm$ 7.04   & $-$3.91$\pm$ 6.40  \\
G165.131$-$13.616 & 0.1581$\pm$0.1387 & 6.07$\pm$1.31  & 5.7$\pm$1.6  & 13.6$\pm$1.6 & $-$1.3$\pm$0.4 & 0.086$\pm$0.115    & $-$1.818$\pm$0.072 & $-$58.3  & $-$36.11$\pm$13.97  & 34.92$\pm$ 6.45   & $-$14.36$\pm$13.05\\
G208.465$-$30.837 & 0.3344$\pm$0.2696 & 2.73$\pm$1.06  & 2.8$\pm$1.2  & 10.4$\pm$1.0 & $-$1.4$\pm$0.6 & 3.736$\pm$0.115    & $-$0.374$\pm$0.094 & 16.3    & $-$35.35$\pm$12.20  & 0.20$\pm$ 4.52    & 26.55$\pm$14.09   \\
\enddata
\tablecomments{Column 1 gives the source name. Columns 2--4 give the corrected Gaia DR3 parallaxes, the mid-IR PLR distances, and error-weighted distances of mid-IR PLR distances and Gaia trigonometric distance, which we adopt in this work. Columns 5 and 6 list the Galactocentric distance and distance to the Galactic plane. Columns 7 and 8 give the proper motion from the Gaia DR3 in the eastward ($\mu_{\rm x}$ = $\mu_{\alpha}$cos$\delta$) and in the northward ($\mu_{\rm y}$ = $\mu_{\delta}$) directions.  Column 9 gives the velocity which is derived from SiO maser detections. For the sources detected in both SiO maser transitions, we take the average of these SiO maser velocities. Columns 10 --12 give the three components of the source peculiar motion toward the Galactic center ($U_{\rm s}$), in the direction of Galactic rotation ($V_{\rm s}$) and toward the north Galactic pole ($W_{\rm s}$), respectively. 
}
\end{deluxetable*}

\subsection{Kinematics}\label{sec:kin}

\begin{figure*}[htbp]
\centering
\includegraphics[width=0.46\textwidth]{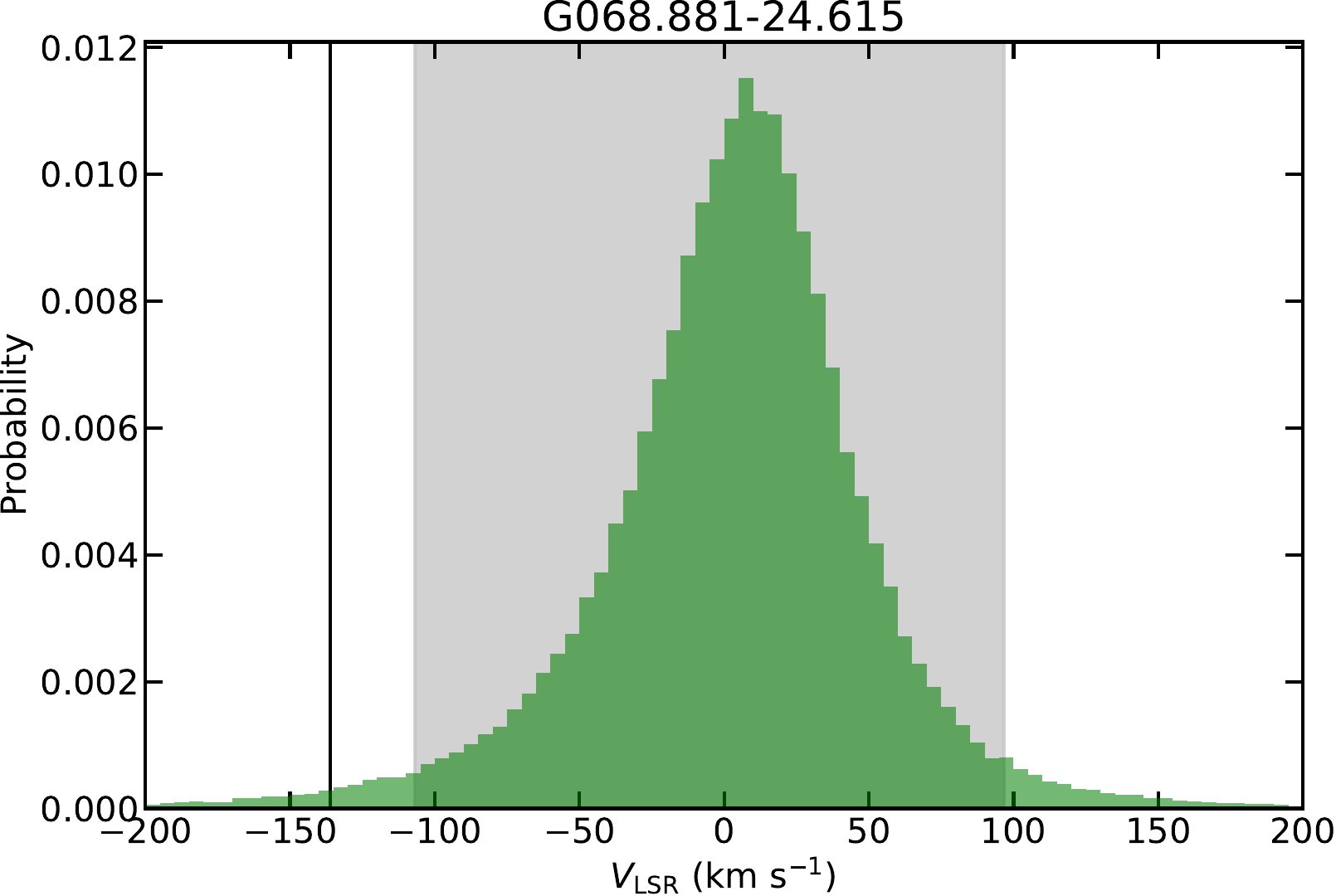}
\includegraphics[width=0.46\textwidth]{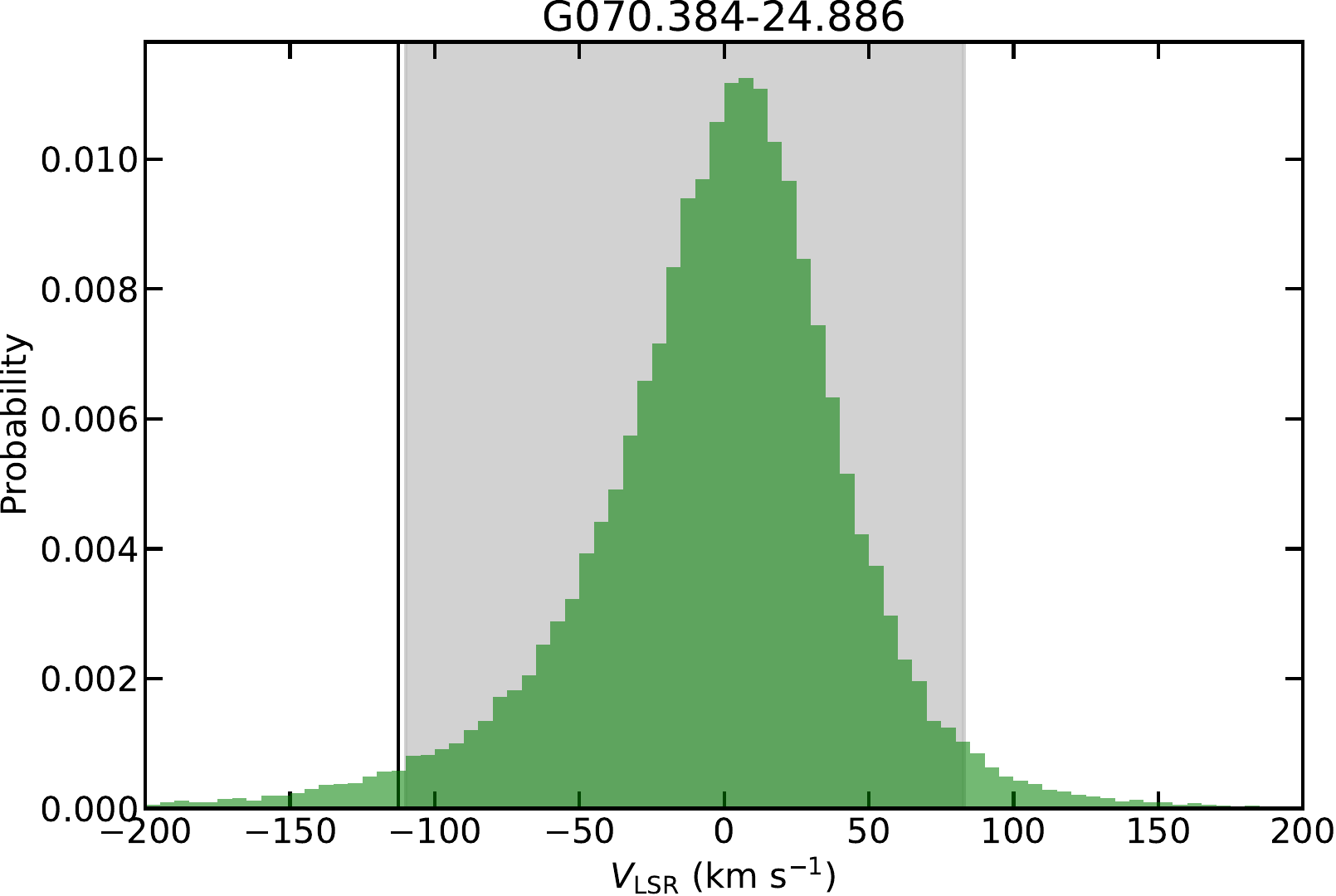}
\includegraphics[width=0.46\textwidth]{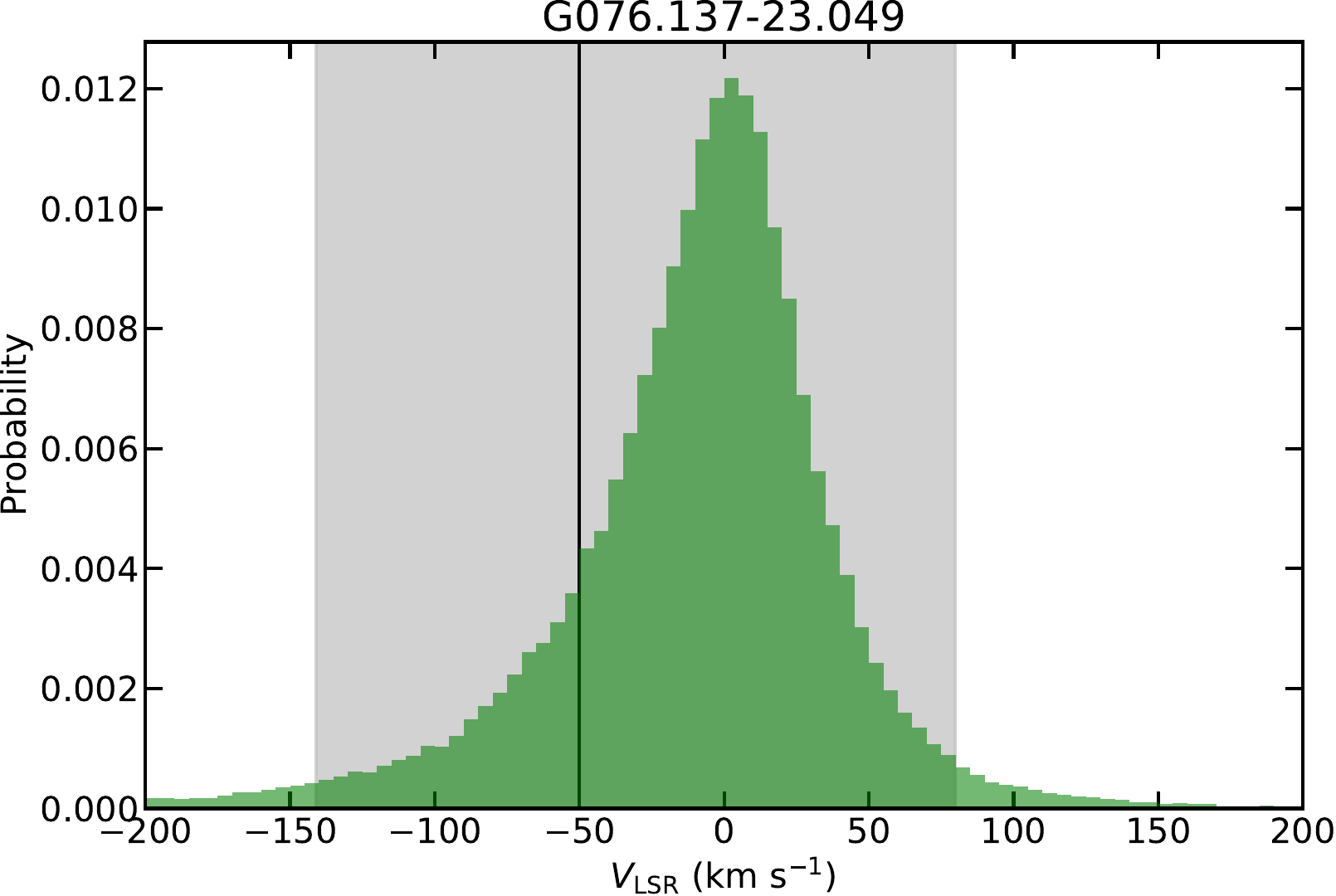}
\includegraphics[width=0.46\textwidth]{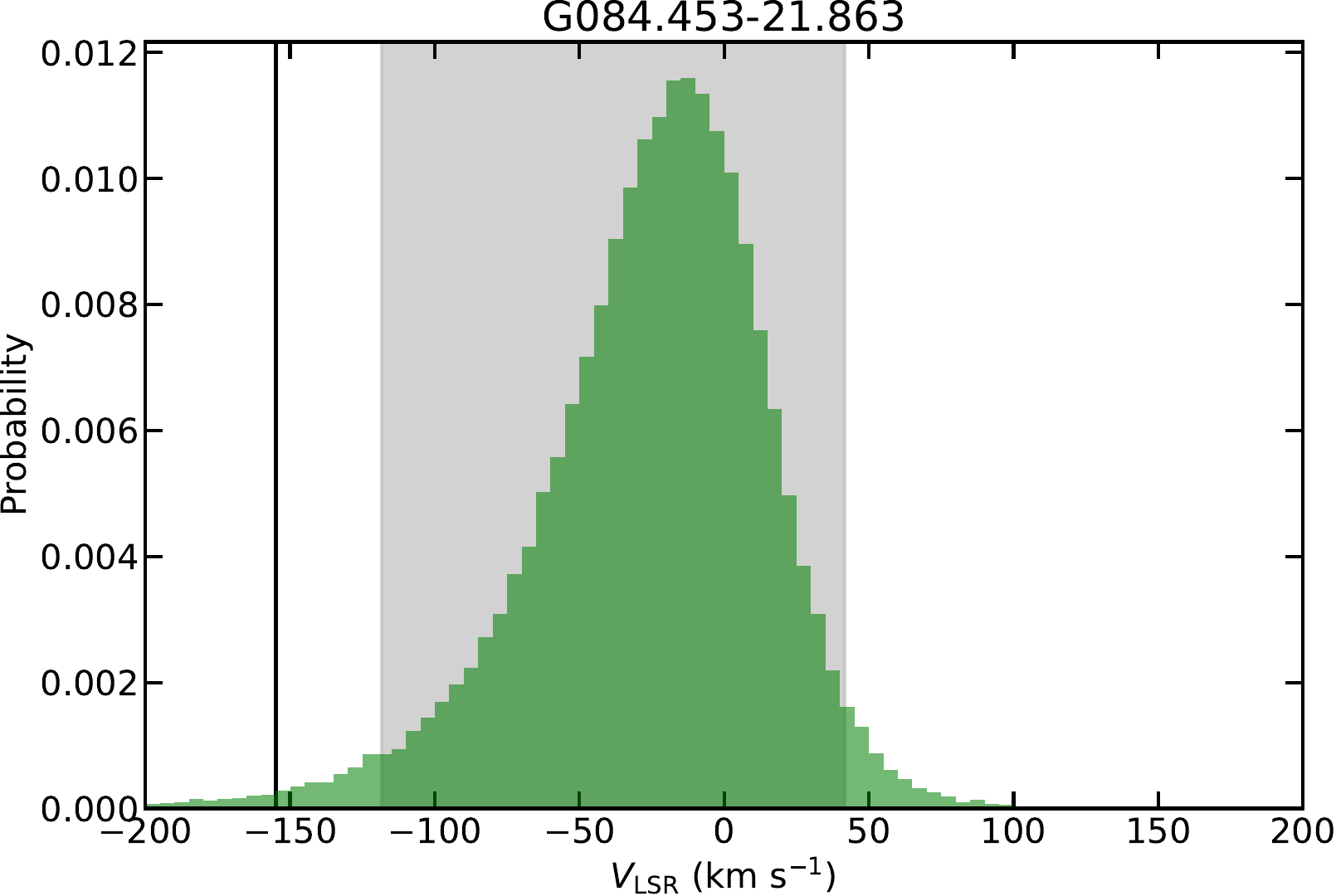}
\includegraphics[width=0.46\textwidth]{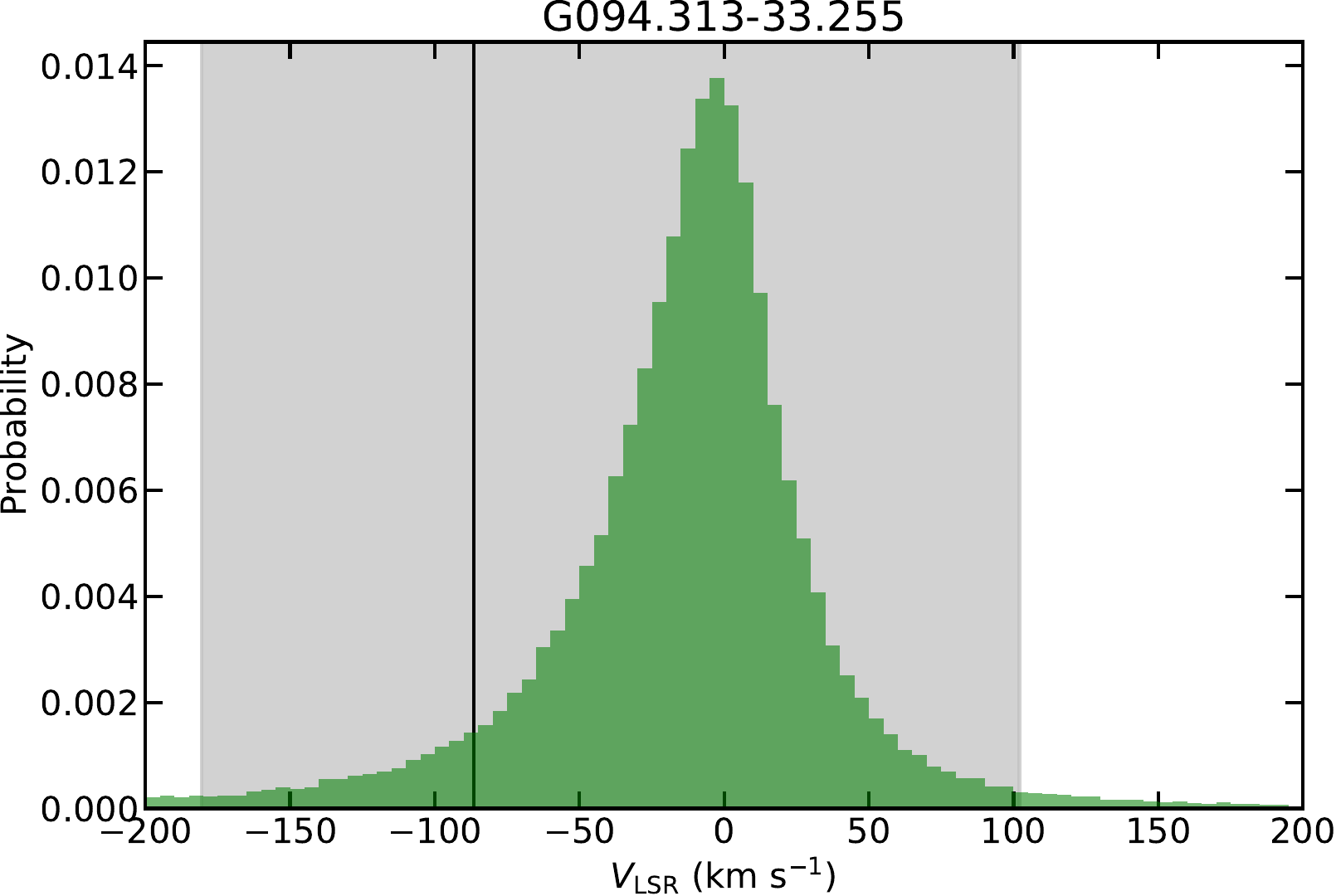}
\includegraphics[width=0.46\textwidth]{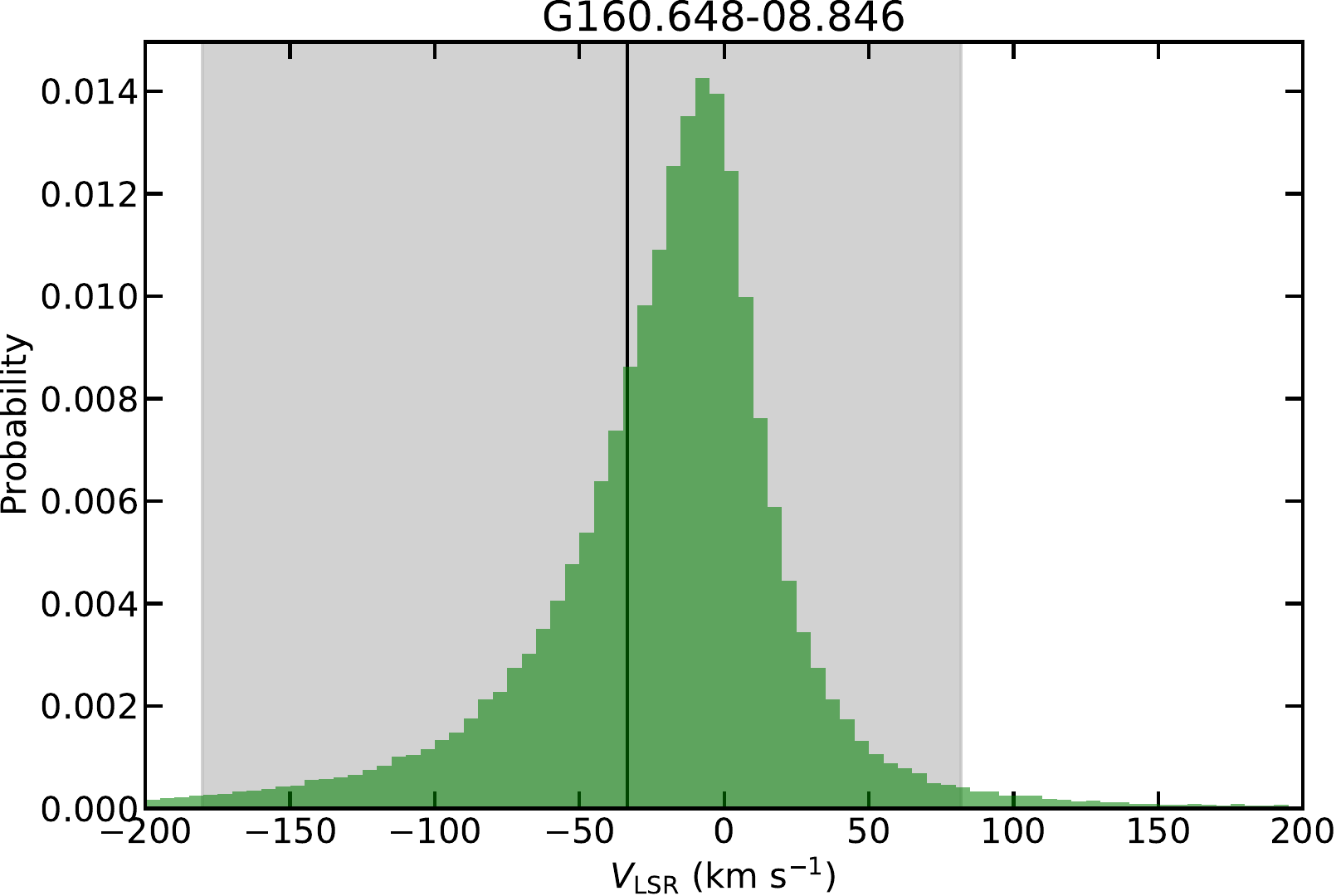}
\includegraphics[width=0.46\textwidth]{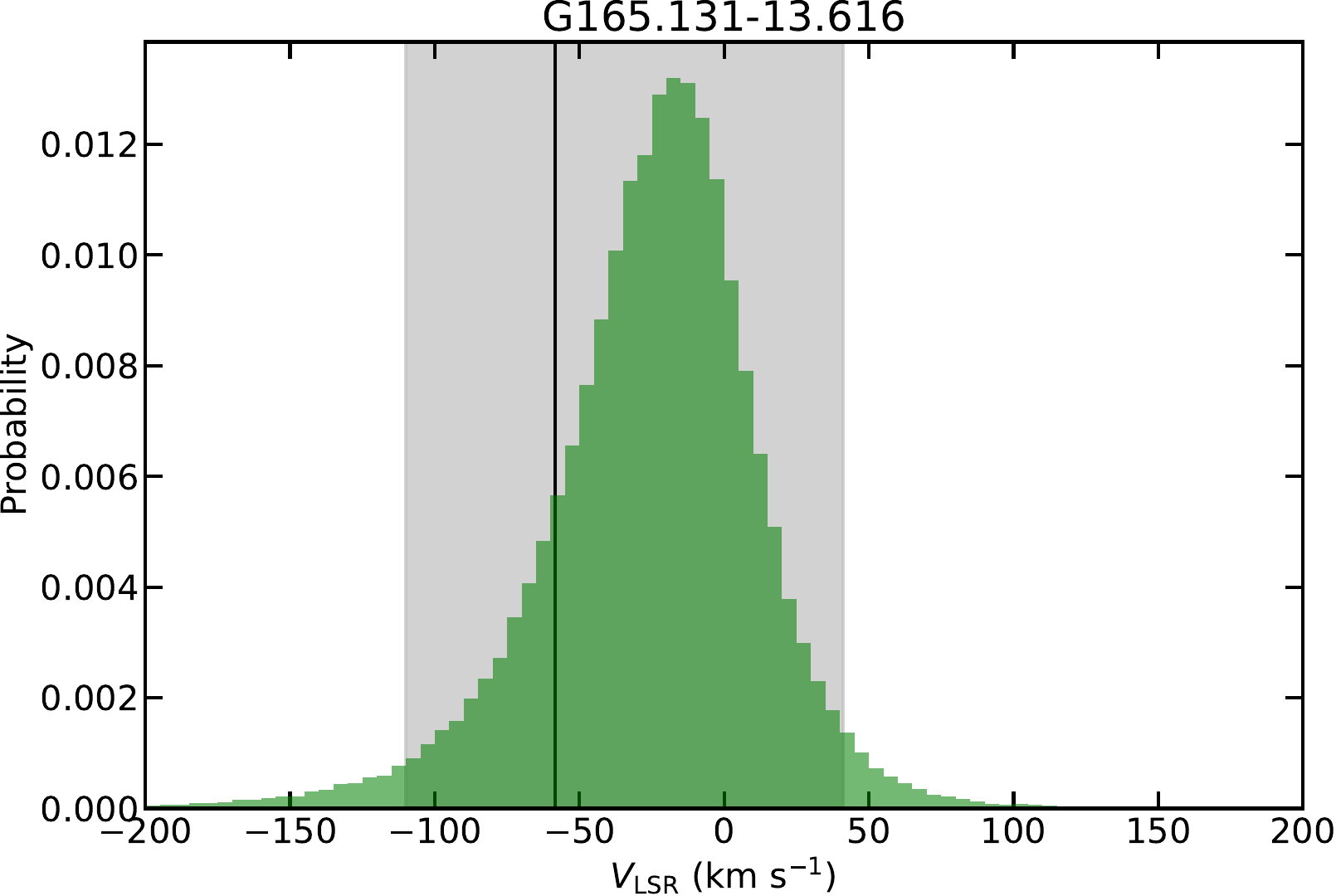}
\includegraphics[width=0.46\textwidth]{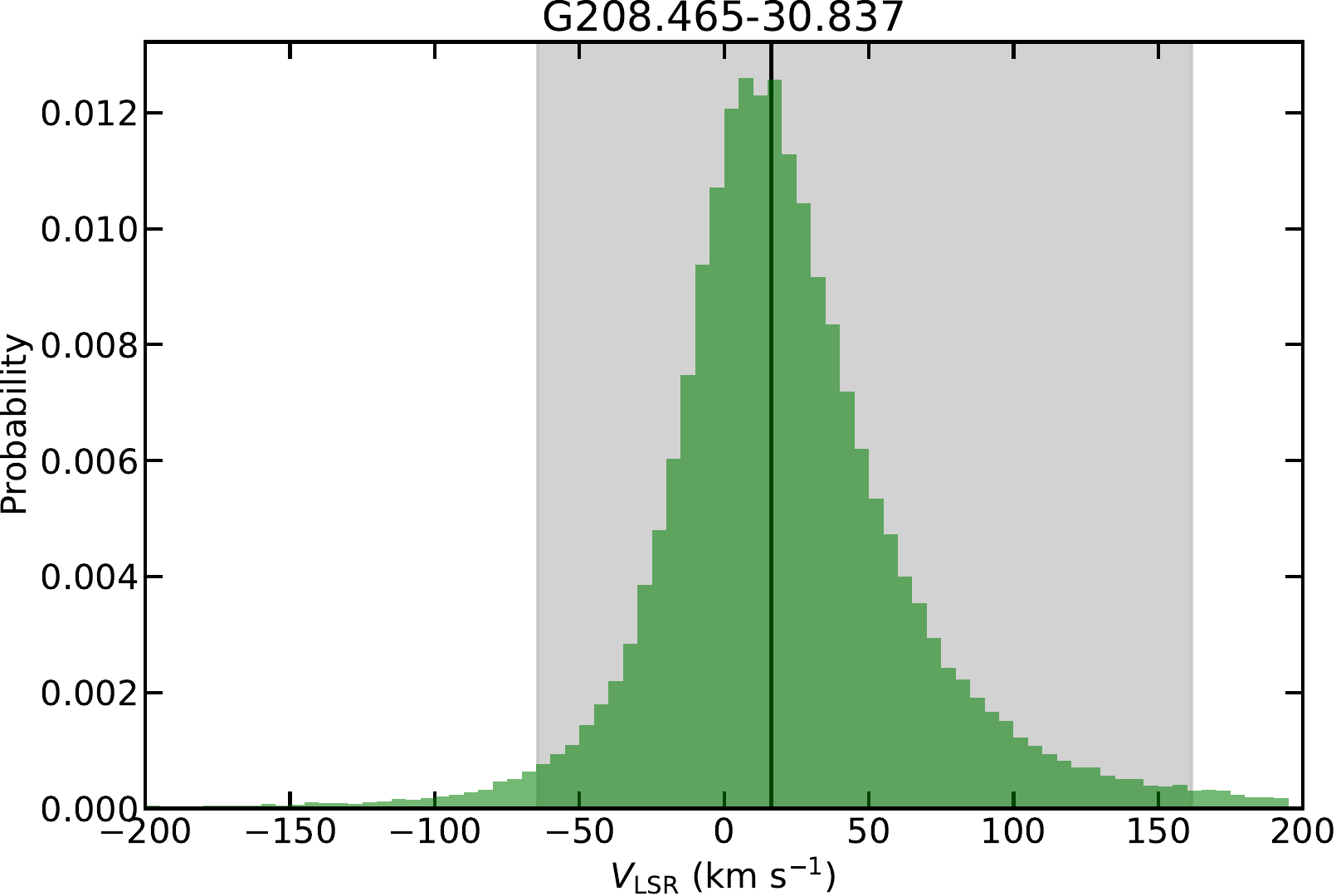}
\caption{Probability density distributions of the LSR velocities for the eight sources. The green histograms represent the simulated distribution from a Monte Carlo analysis. The vertical solid lines represent the observed stellar LSR velocities based on SiO maser spectra. 
The shaded region corresponds to the 95.45\% confidence level (2$\sigma$).
\label{fig:velo-simulation}}
\end{figure*}

Utilizing the Galactic coordinates, distances, and a Galactic rotation model, we can examine if the observed LSR velocities can be explained by the Galactic circular motion. We adopt the best-fit model rotation curve for the Galaxy (A5) discussed by \citet{2019ApJ...885..131R}. This model is characterized by a flat Galactic rotation curve with the distance from the Sun to the Galactic center $R_0$ of 8.15~$\pm$~0.15~kpc and a Galactic rotation speed near the solar circle $\Theta_0$ of 236~$\pm$~7~km~s$^{-1}$.
Assuming a velocity dispersion of 30~\kms\,caused by peculiar motions in each of the three dimensions for the observed stars in Galactocentric Cartesian coordinates (i.e., X, Y, Z in Fig.~\ref{fig:3dplot}) and taking the errors in the adopted parameters into account,  we can evaluate the probability of different LSR velocities by applying a Monte Carlo approach where 100 000 simulations are performed. The probability distribution of the LSR velocity for the eight maser-host stars is shown in Fig.~\ref{fig:velo-simulation}. Based on this plot, we find that the observed SiO maser velocities of G068.881$-$24.615, G070.384$-$24.886 and G084.453$-$21.863 deviate from the expected LSR velocity distribution caused by the circular Galactic rotation at the 95.45\% confidence level (2$\sigma$). This suggests that the motions of the three stars are unlikely to follow the Galactic rotation.

With coordinates, distances, proper motions, and radial velocities, we are also able to determine the 3D velocities of our stars. 
The 3D motions are characterized by a circular rotation component and non-circular (i.e., peculiar) velocity components, $U_s$, $V_s$, $W_s$, oriented toward the Galactic Center, in the direction of Galactic rotation, and toward the north Galactic pole, respectively. 
Again, we adopt the A5 model \citep{2019ApJ...885..131R}, and then follow the method described by \citet{2009ApJ...700..137R} to calculate the non-circular velocity in the Galactocentric reference frame. Table~\ref{Tab:kin} lists the non-circular velocity components of our sources.
The two-dimensional projections of the peculiar motions in the Galactocentric Cartesian frame are shown as arrows in Fig.~\ref{fig:3dplot}.
We found that all sources have negative $U_s$, which indicates that they are moving away from the Galactic center. 
Halo stars are usually defined as stars with $v_{\rm tot}$ = $\sqrt[]{U_s^2+V_s^2+W_s^2} >$ 180~\kms\,\citep[e.g.,][]{2004AJ....128.1177V,2010A&A...511L..10N}. Given this, G068.881$-$24.615, with $v_{\rm tot} >$180~\kms, is the only star with a possible location in the Galactic halo.

Our target stars were initially proposed to possibly belong to the Sgr stream \citep{Mauron2019}. This scenario can be verified or disproved based on their motions. We calculate the angles, $\phi_{\rm Sgr}$, between the 3D velocities ($U_s$, $V_s$+$\Theta_0$, $W_s$) of the eight sources and the Sgr orbital plane, as well as the angles ($\phi_{\rm Gp}$) between non-circular velocities and the Galactic plane. We find that the values of $\phi_{\rm Sgr}$ are all larger than 60$^\circ$, while $\phi_{\rm Gp}$ are all smaller than 10$^\circ$. If our stars belong to the Sgr stream, their motions should align with the Sgr orbital plane (i.e., $\phi_{\rm Sgr}<$30$^\circ$). Their large $\phi_{\rm Sgr}$ values further suggest that these SiO maser-traced O-rich AGBs are unlikely to arise from the Sgr stream. 

We caution that the rotation speed of the Galactic thick disk can be slower than that of the thin disk (see Table 3 in \citealt{2004AJ....128.1177V} for example). The peculiar motions of our maser-host stars are calculated based on a rotation curve that is derived from the thin disk objects, namely (young) regions of high-mass star formation. Hence, we may underestimate the value of $V_s$ and $v_{\rm tot}$ for these stars, which implies that the peculiar motions can be even larger than the values presented above if we adopt a slower  rotation speed. Therefore, our conclusions still hold. 



\section{Summary}\label{Sec:sum}
We carried out a sensitive SiO maser ($J$=1--0, $v$=1, 2) survey toward a sample of 102 off-plane O-rich AGBs selected from \citet{Mauron2019}, using the Effelsberg-100 m and Tianma-65m telescopes. 
The main results can be summarized as follows:

\begin{itemize}
\item[1.] SiO masers are newly detected toward eight O-rich AGB stars, which corresponds to a detection rate of 8$\pm$3\%. The maser velocities provide 
the stellar radial velocities for the maser-host stars for the first time. The radial velocities of three stars (i.e., G068.881$-$24.615, G070.384$-$24.886, and G084.453$-$21.863) significantly deviate from the values expected from Galactic circular motion. \\

\item[2.] We revisit the distances of the maser-host stars and found that distances, based on a near-IR period-luminosity relation, given in \cite{Mauron2019} could be overestimated by 60\%. Except for G160.648$-$08.846, the other stars are significantly below the Galactic plane with $|Z|=$ 1.3--2.3~kpc.\\

\item[3.] With the updated Galactic distribution, stellar velocity, and proper motion information from Gaia DR3, we derived the 3D motions of the eight maser-host AGBs, and found that (i) they are all moving away from the Galactic Center, and (ii) G068.881$-$24.615 is likely to arise from the Galactic halo, while G160.648$-$08.846 is probably located in the Galactic thin disk, and the other six stars probably are within the Galaxy's thick disk.\\

\end{itemize}


In future work, extending the SiO maser survey to higher $|Z|$ is necessary for laying the foundation to reveal the structure and nature of the Galactic thick disk, halo, and even the tidal stellar streams. 
In addition, follow-up VLBI observations toward these masers have the great potential to measure their parallaxes and proper motions with much higher accuracy, which could eventually pave the way toward understanding the dynamic evolution of the Milky Way. Future acquisition of information on the metallicity of these stars can also be used to confirm their nature and origins.

\begin{acknowledgments}
We thank the anonymous referee for constructive comments that improved this paper.
We appreciate Dr. Xunchuan Liu for providing useful information about the Q-band observations of Tianma telescope.
We thank staffs in the Effelsberg-100 m and Tianma telescopes for their assistance with our observations. 
We acknowledge the support from NSFC grant under No. 12273010.
NM thanks the Director of LUPM, Denis Puy, for hospitality in the laboratory and encouragement.
This work is based on observations with the 100-m telescope of the MPIfR (Max-Plank-Institut f$\ddot{u}$r Radioastronomie) at Effelsberg. The user interface of the luminosity calculator is available from \href{https://github.com/gongyan2444/data/blob/master/luminosity.py}{this link}. 
\end{acknowledgments}

%

\vspace{5mm}
\facilities{Effelsberg-100 m, Tianma-65 m, Gaia}


\software{GILDAS, python}



\appendix
In this Appendix, we provide the source information and 1$\sigma$ noise level of the 94 non-detections in our survey. These details are listed in Table~\ref{Tab:non-det}.

\begin{deluxetable}{lccccccc}[htbp]
\tablecaption{Observed source information and 1$\sigma$ noise level of both SiO maser transitions for non-detections.\label{Tab:non-det}}  
\centering
\tabletypesize{\small}
\tablewidth{700 pt}
\tablecolumns{8}
\tablenum{A.1}
\renewcommand\arraystretch{1}
\centering
\setlength{\tabcolsep}{10pt}
\tablehead{
\colhead{Name} & \colhead{$\alpha_{\rm J2000}$} & \colhead{$\delta_{\rm J2000}$} & \colhead{$V_{\rm radial}$} & \colhead{Corrected $K_s$} & \colhead{Period}  &  \colhead{$\sigma_{v=1}$} & \colhead{$\sigma_{v=2}$} \\
\colhead{} & \colhead{} & \colhead{}  & \colhead{(\kms)} & \colhead{(mag)} & \colhead{(day)} & \colhead{(Jy)} & \colhead{(Jy)} } 
\startdata
\multicolumn{8}{c}{Effelsberg-100 m observations} \\
\hline
G057.196$-$38.104 & 21:48:04.87 & $+$00:22:02.1 & \nodata  & 7.129  & 214.6 & 0.05 & 0.05 \\
G067.572$-$24.040 & 21:25:44.07 & $+$16:02:10.9 & \nodata  & 6.668  & 220.8 & 0.06 & 0.06 \\
G072.159$-$38.360 & 22:19:34.60 & $+$09:06:37.5 & \nodata  & 6.424  & 230.9 & 0.04 & 0.05 \\
G072.398$-$42.331 & 22:31:14.65 & $+$06:22:49.1 & \nodata  & 6.744  & 222.9 & 0.04 & 0.04 \\
G072.440$-$20.303 & 21:26:27.68 & $+$21:53:18.6 & \nodata  & 8.761  & 202.8 & 0.04 & 0.05 \\
\hline
\multicolumn{8}{c}{Tianma-65 m observations} \\
\hline
G150.030$+$19.764 & 06:08:36.06 &  $+$64:07:57.8  & \nodata  & 8.721  & 261.3  & 0.04$\dagger$ & 0.03 \\
G161.049$+$21.323 & 06:47:13.83 &  $+$54:53:33.2  & $-$91.6  & 7.275  & 253.9  & 0.03 & 0.02 \\
G161.414$+$32.299 & 08:04:35.51 &  $+$56:21:35.7  & \nodata  & 5.751  & 342.5  & 0.03$\dagger$ & 0.03 \\
G161.629$+$15.817 & 06:13:45.06 &  $+$52:25:40.8  & \nodata  & 7.208  & 265.2  & 0.03 & 0.03 \\
G162.618$+$18.961 & 06:35:16.14 &  $+$52:44:23.6  & \nodata  & 5.806  & 259.6  & 0.03 & 0.03 \\
\enddata
\tablecomments{Columns 1--3 give the source name and coordinates, arranged by increasing Galactic longitude. Column 4 lists the radial velocities of the 49 sources provided by the Gaia DR3 catalog \citep{2023gaiadr3}.
Column 5--6 give the corrected $K_s$ magnitude and period which are adopted from \cite{Mauron2019}.
Columns 7--8 list the 1$\sigma$ noise level of SiO $v = 1$ and $v = 2$, respectively.
The marker ``$\dagger$" in Column 7 indicates that only left circular polarization data are available for this source, due to instrumental reasons. \\
The first five sources of each survey are shown here for guidance, the full table is available online.}
\end{deluxetable}


\bibliography{reference}{}

\begin{thebibliography}{}
\expandafter\ifx\csname natexlab\endcsname\relax\def\natexlab#1{#1}\fi
\providecommand{\url}[1]{\href{#1}{#1}}
\providecommand{\dodoi}[1]{doi:~\href{http://doi.org/#1}{\nolinkurl{#1}}}
\providecommand{\doeprint}[1]{\href{http://ascl.net/#1}{\nolinkurl{http://ascl.net/#1}}}
\providecommand{\doarXiv}[1]{\href{https://arxiv.org/abs/#1}{\nolinkurl{https://arxiv.org/abs/#1}}}

\bibitem[{{Abdurro'uf} {et~al.}(2022){Abdurro'uf}, {Accetta}, {Aerts}, {Silva
  Aguirre}, {Ahumada}, {Ajgaonkar}, {Filiz Ak}, {Alam}, {Allende Prieto},
  {Almeida}, {Anders}, {Anderson}, {Andrews}, {Anguiano}, {Aquino-Ort{\'\i}z},
  {Arag{\'o}n-Salamanca}, {Argudo-Fern{\'a}ndez}, {Ata}, {Aubert},
  {Avila-Reese}, {Badenes}, {Barb{\'a}}, {Barger}, {Barrera-Ballesteros},
  {Beaton}, {Beers}, {Belfiore}, {Bender}, {Bernardi}, {Bershady}, {Beutler},
  {Bidin}, {Bird}, {Bizyaev}, {Blanc}, {Blanton}, {Boardman}, {Bolton},
  {Boquien}, {Borissova}, {Bovy}, {Brandt}, {Brown}, {Brownstein}, {Brusa},
  {Buchner}, {Bundy}, {Burchett}, {Bureau}, {Burgasser}, {Cabang}, {Campbell},
  {Cappellari}, {Carlberg}, {Wanderley}, {Carrera}, {Cash}, {Chen}, {Chen},
  {Cherinka}, {Chiappini}, {Choi}, {Chojnowski}, {Chung}, {Clerc}, {Cohen},
  {Comerford}, {Comparat}, {da Costa}, {Covey}, {Crane}, {Cruz-Gonzalez},
  {Culhane}, {Cunha}, {Dai}, {Damke}, {Darling}, {Davidson}, {Davies},
  {Dawson}, {De Lee}, {Diamond-Stanic}, {Cano-D{\'\i}az}, {S{\'a}nchez},
  {Donor}, {Duckworth}, {Dwelly}, {Eisenstein}, {Elsworth}, {Emsellem},
  {Eracleous}, {Escoffier}, {Fan}, {Farr}, {Feng}, {Fern{\'a}ndez-Trincado},
  {Feuillet}, {Filipp}, {Fillingham}, {Frinchaboy}, {Fromenteau}, {Galbany},
  {Garc{\'\i}a}, {Garc{\'\i}a-Hern{\'a}ndez}, {Ge}, {Geisler}, {Gelfand},
  {G{\'e}ron}, {Gibson}, {Goddy}, {Godoy-Rivera}, {Grabowski}, {Green},
  {Greener}, {Grier}, {Griffith}, {Guo}, {Guy}, {Hadjara}, {Harding},
  {Hasselquist}, {Hayes}, {Hearty}, {Hern{\'a}ndez}, {Hill}, {Hogg},
  {Holtzman}, {Horta}, {Hsieh}, {Hsu}, {Hsu}, {Huber}, {Huertas-Company},
  {Hutchinson}, {Hwang}, {Ibarra-Medel}, {Chitham}, {Ilha}, {Imig}, {Jaekle},
  {Jayasinghe}, {Ji}, {Johnson}, {Jones}, {J{\"o}nsson}, {Katkov}, {Khalatyan},
  {Kinemuchi}, {Kisku}, {Knapen}, {Kneib}, {Kollmeier}, {Kong}, {Kounkel},
  {Kreckel}, {Krishnarao}, {Lacerna}, {Lane}, {Langgin}, {Lavender}, {Law},
  {Lazarz}, {Leung}, {Leung}, {Lewis}, {Li}, {Li}, {Lian}, {Liang}, {Lin},
  {Lin}, {Lin}, {Lintott}, {Long}, {Longa-Pe{\~n}a}, {L{\'o}pez-Cob{\'a}},
  {Lu}, {Lundgren}, {Luo}, {Mackereth}, {de la Macorra}, {Mahadevan},
  {Majewski}, {Manchado}, {Mandeville}, {Maraston}, {Margalef-Bentabol},
  {Masseron}, {Masters}, {Mathur}, {McDermid}, {Mckay}, {Merloni},
  {Merrifield}, {Meszaros}, {Miglio}, {Di Mille}, {Minniti}, {Minsley},
  {Monachesi}, {Moon}, {Mosser}, {Mulchaey}, {Muna}, {Mu{\~n}oz}, {Myers},
  {Myers}, {Nadathur}, {Nair}, {Nandra}, {Neumann}, {Newman}, {Nidever},
  {Nikakhtar}, {Nitschelm}, {O'Connell}, {Garma-Oehmichen}, {Luan Souza de
  Oliveira}, {Olney}, {Oravetz}, {Ortigoza-Urdaneta}, {Osorio}, {Otter},
  {Pace}, {Padilla}, {Pan}, {Pan}, {Parikh}, {Parker}, {Peirani}, {Pe{\~n}a
  Ram{\'\i}rez}, {Penny}, {Percival}, {Perez-Fournon}, {Pinsonneault},
  {Poidevin}, {Poovelil}, {Price-Whelan}, {B{\'a}rbara de Andrade Queiroz},
  {Raddick}, {Ray}, {Rembold}, {Riddle}, {Riffel}, {Riffel}, {Rix}, {Robin},
  {Rodr{\'\i}guez-Puebla}, {Roman-Lopes}, {Rom{\'a}n-Z{\'u}{\~n}iga}, {Rose},
  {Ross}, {Rossi}, {Rubin}, {Salvato}, {S{\'a}nchez}, {S{\'a}nchez-Gallego},
  {Sanderson}, {Santana Rojas}, {Sarceno}, {Sarmiento}, {Sayres}, {Sazonova},
  {Schaefer}, {Schiavon}, {Schlegel}, {Schneider}, {Schultheis}, {Schwope},
  {Serenelli}, {Serna}, {Shao}, {Shapiro}, {Sharma}, {Shen}, {Shetrone}, {Shu},
  {Simon}, {Skrutskie}, {Smethurst}, {Smith}, {Sobeck}, {Spoo}, {Sprague},
  {Stark}, {Stassun}, {Steinmetz}, {Stello}, {Stone-Martinez},
  {Storchi-Bergmann}, {Stringfellow}, {Stutz}, {Su}, {Taghizadeh-Popp},
  {Talbot}, {Tayar}, {Telles}, {Teske}, {Thakar}, {Theissen}, {Tkachenko},
  {Thomas}, {Tojeiro}, {Hernandez Toledo}, {Troup}, {Trump}, {Trussler},
  {Turner}, {Tuttle}, {Unda-Sanzana}, {V{\'a}zquez-Mata}, {Valentini},
  {Valenzuela}, {Vargas-Gonz{\'a}lez}, {Vargas-Maga{\~n}a}, {Alfaro},
  {Villanova}, {Vincenzo}, {Wake}, {Warfield}, {Washington}, {Weaver},
  {Weijmans}, {Weinberg}, {Weiss}, {Westfall}, {Wild}, {Wilde}, {Wilson},
  {Wilson}, {Wilson}, {Wolf}, {Wood-Vasey}, {Yan}, {Zamora}, {Zasowski},
  {Zhang}, {Zhao}, {Zheng}, {Zheng}, \& {Zhu}}]{2022ApJS..259...35A}
{Abdurro'uf}, {Accetta}, K., {Aerts}, C., {et~al.} 2022, \apjs, 259, 35,
  \dodoi{10.3847/1538-4365/ac4414}

\bibitem[{{Andriantsaralaza} {et~al.}(2022){Andriantsaralaza}, {Ramstedt},
  {Vlemmings}, \& {De Beck}}]{2022A&A...667A..74A}
{Andriantsaralaza}, M., {Ramstedt}, S., {Vlemmings}, W.~H.~T., \& {De Beck}, E.
  2022, \aap, 667, A74, \dodoi{10.1051/0004-6361/202243670}

\bibitem[{{Bland-Hawthorn} \& {Gerhard}(2016)}]{2016ARA&A..54..529B}
{Bland-Hawthorn}, J., \& {Gerhard}, O. 2016, \araa, 54, 529,
  \dodoi{10.1146/annurev-astro-081915-023441}

\bibitem[{{Bussa} \& {VEGAS Development Team}(2012)}]{2012AAS...21944610B}
{Bussa}, S., \& {VEGAS Development Team}. 2012, in American Astronomical
  Society Meeting Abstracts 219, 446.10

\bibitem[{{de Jong} {et~al.}(2010){de Jong}, {Yanny}, {Rix}, {Dolphin},
  {Martin}, \& {Beers}}]{2010ApJ...714..663D}
{de Jong}, J. T.~A., {Yanny}, B., {Rix}, H.-W., {et~al.} 2010, \apj, 714, 663,
  \dodoi{10.1088/0004-637X/714/1/663}

\bibitem[{{Deguchi} {et~al.}(2007{\natexlab{a}}){Deguchi}, {Nakashima}, {Kwok},
  \& {Koning}}]{2007ApJ...664.1130D}
{Deguchi}, S., {Nakashima}, J.-i., {Kwok}, S., \& {Koning}, N.
  2007{\natexlab{a}}, \apj, 664, 1130, \dodoi{10.1086/519154}

\bibitem[{{Deguchi} {et~al.}(2010){Deguchi}, {Shimoikura}, \&
  {Koike}}]{2010PASJ...62..525D}
{Deguchi}, S., {Shimoikura}, T., \& {Koike}, K. 2010, \pasj, 62, 525,
  \dodoi{10.1093/pasj/62.3.525}

\bibitem[{{Deguchi} {et~al.}(2004){Deguchi}, {Fujii}, {Glass}, {Imai}, {Ita},
  {Izumiura}, {Kameya}, {Miyazaki}, {Nakada}, \&
  {Nakashima}}]{2004PASJ...56..765D}
{Deguchi}, S., {Fujii}, T., {Glass}, I.~S., {et~al.} 2004, \pasj, 56, 765,
  \dodoi{10.1093/pasj/56.5.765}

\bibitem[{{Deguchi} {et~al.}(2007{\natexlab{b}}){Deguchi}, {Fujii}, {Ita},
  {Imai}, {Izumiura}, {Kameya}, {Matsunaga}, {Miyazaki}, {Mizutani}, {Nakada},
  {Nakashima}, \& {Winnberg}}]{2007PASJ...59..559D}
{Deguchi}, S., {Fujii}, T., {Ita}, Y., {et~al.} 2007{\natexlab{b}}, \pasj, 59,
  559, \dodoi{10.1093/pasj/59.3.559}

\bibitem[{{Desmurs} {et~al.}(2014){Desmurs}, {Bujarrabal}, {Lindqvist},
  {Alcolea}, {Soria-Ruiz}, \& {Bergman}}]{2014A&A...565A.127D}
{Desmurs}, J.~F., {Bujarrabal}, V., {Lindqvist}, M., {et~al.} 2014, \aap, 565,
  A127, \dodoi{10.1051/0004-6361/201423550}

\bibitem[{{Dong} {et~al.}(2018){Dong}, {Zhong}, {Wang}, {Liu}, \&
  {Shen}}]{2018ITAP...66.2044D}
{Dong}, J., {Zhong}, W., {Wang}, J., {Liu}, Q., \& {Shen}, Z. 2018, \itap, 66,
  2044, \dodoi{10.1109/TAP.2018.2796378}

\bibitem[{{Drake} {et~al.}(2014){Drake}, {Graham}, {Djorgovski}, {Catelan},
  {Mahabal}, {Torrealba}, {Garc{\'\i}a-{\'A}lvarez}, {Donalek}, {Prieto},
  {Williams}, {Larson}, {Christen sen}, {Belokurov}, {Koposov}, {Beshore},
  {Boattini}, {Gibbs}, {Hill}, {Kowalski}, {Johnson}, \&
  {Shelly}}]{2014ApJS..213....9D}
{Drake}, A.~J., {Graham}, M.~J., {Djorgovski}, S.~G., {et~al.} 2014, \apjs,
  213, 9, \dodoi{10.1088/0067-0049/213/1/9}

\bibitem[{{Eder} {et~al.}(1988){Eder}, {Lewis}, \&
  {Terzian}}]{1988ApJS...66..183E}
{Eder}, J., {Lewis}, B.~M., \& {Terzian}, Y. 1988, \apjs, 66, 183,
  \dodoi{10.1086/191252}

\bibitem[{{Engels} \& {Lewis}(1996)}]{1996A&AS..116..117E}
{Engels}, D., \& {Lewis}, B.~M. 1996, \aaps, 116, 117

\bibitem[{{Fabricius} {et~al.}(2021){Fabricius}, {Luri}, {Arenou}, {Babusiaux},
  {Helmi}, {Muraveva}, {Reyl{\'e}}, {Spoto}, {Vallenari}, {Antoja}, {Balbinot},
  {Barache}, {Bauchet}, {Bragaglia}, {Busonero}, {Cantat-Gaudin}, {Carrasco},
  {Diakit{\'e}}, {Fabrizio}, {Figueras}, {Garcia-Gutierrez}, {Garofalo},
  {Jordi}, {Kervella}, {Khanna}, {Leclerc}, {Licata}, {Lambert}, {Marrese},
  {Masip}, {Ramos}, {Robichon}, {Robin}, {Romero-G{\'o}mez}, {Rubele}, \&
  {Weiler}}]{2021A&A...649A...5F}
{Fabricius}, C., {Luri}, X., {Arenou}, F., {et~al.} 2021, \aap, 649, A5,
  \dodoi{10.1051/0004-6361/202039834}

\bibitem[{{Fonfr{\'\i}a Exp{\'o}sito} {et~al.}(2006){Fonfr{\'\i}a
  Exp{\'o}sito}, {Ag{\'u}ndez}, {Tercero}, {Pardo}, \&
  {Cernicharo}}]{2006ApJ...646L.127F}
{Fonfr{\'\i}a Exp{\'o}sito}, J.~P., {Ag{\'u}ndez}, M., {Tercero}, B., {Pardo},
  J.~R., \& {Cernicharo}, J. 2006, \apjl, 646, L127, \dodoi{10.1086/507104}

\bibitem[{{Gaia Collaboration} {et~al.}(2023){Gaia Collaboration}, {Vallenari},
  {Brown}, {Prusti}, {de Bruijne}, {Arenou}, {Babusiaux}, {Biermann},
  {Creevey}, {Ducourant}, {Evans}, {Eyer}, {Guerra}, {Hutton}, {Jordi},
  {Klioner}, {Lammers}, {Lindegren}, {Luri}, {Mignard}, {Panem}, {Pourbaix},
  {Randich}, {Sartoretti}, {Soubiran}, {Tanga}, {Walton}, {Bailer-Jones},
  {Bastian}, {Drimmel}, {Jansen}, {Katz}, {Lattanzi}, {van Leeuwen}, {Bakker},
  {Cacciari}, {Casta{\~n}eda}, {De Angeli}, {Fabricius}, {Fouesneau},
  {Fr{\'e}mat}, {Galluccio}, {Guerrier}, {Heiter}, {Masana}, {Messineo},
  {Mowlavi}, {Nicolas}, {Nienartowicz}, {Pailler}, {Panuzzo}, {Riclet}, {Roux},
  {Seabroke}, {Sordo}, {Th{\'e}venin}, {Gracia-Abril}, {Portell}, {Teyssier},
  {Altmann}, {Andrae}, {Audard}, {Bellas-Velidis}, {Benson}, {Berthier},
  {Blomme}, {Burgess}, {Busonero}, {Busso}, {C{\'a}novas}, {Carry}, {Cellino},
  {Cheek}, {Clementini}, {Damerdji}, {Davidson}, {de Teodoro}, {Nu{\~n}ez
  Campos}, {Delchambre}, {Dell'Oro}, {Esquej}, {Fern{\'a}ndez-Hern{\'a}ndez},
  {Fraile}, {Garabato}, {Garc{\'\i}a-Lario}, {Gosset}, {Haigron}, {Halbwachs},
  {Hambly}, {Harrison}, {Hern{\'a}ndez}, {Hestroffer}, {Hodgkin}, {Holl},
  {Jan{\ss}en}, {Jevardat de Fombelle}, {Jordan}, {Krone-Martins}, {Lanzafame},
  {L{\"o}ffler}, {Marchal}, {Marrese}, {Moitinho}, {Muinonen}, {Osborne},
  {Pancino}, {Pauwels}, {Recio-Blanco}, {Reyl{\'e}}, {Riello}, {Rimoldini},
  {Roegiers}, {Rybizki}, {Sarro}, {Siopis}, {Smith}, {Sozzetti}, {Utrilla},
  {van Leeuwen}, {Abbas}, {{\'A}brah{\'a}m}, {Abreu Aramburu}, {Aerts},
  {Aguado}, {Ajaj}, {Aldea-Montero}, {Altavilla}, {{\'A}lvarez}, {Alves},
  {Anders}, {Anderson}, {Anglada Varela}, {Antoja}, {Baines}, {Baker},
  {Balaguer-N{\'u}{\~n}ez}, {Balbinot}, {Balog}, {Barache}, {Barbato},
  {Barros}, {Barstow}, {Bartolom{\'e}}, {Bassilana}, {Bauchet}, {Becciani},
  {Bellazzini}, {Berihuete}, {Bernet}, {Bertone}, {Bianchi}, {Binnenfeld},
  {Blanco-Cuaresma}, {Blazere}, {Boch}, {Bombrun}, {Bossini}, {Bouquillon},
  {Bragaglia}, {Bramante}, {Breedt}, {Bressan}, {Brouillet}, {Brugaletta},
  {Bucciarelli}, {Burlacu}, {Butkevich}, {Buzzi}, {Caffau}, {Cancelliere},
  {Cantat-Gaudin}, {Carballo}, {Carlucci}, {Carnerero}, {Carrasco},
  {Casamiquela}, {Castellani}, {Castro-Ginard}, {Chaoul}, {Charlot}, {Chemin},
  {Chiaramida}, {Chiavassa}, {Chornay}, {Comoretto}, {Contursi}, {Cooper},
  {Cornez}, {Cowell}, {Crifo}, {Cropper}, {Crosta}, {Crowley}, {Dafonte},
  {Dapergolas}, {David}, {David}, {de Laverny}, {De Luise}, {De March}, {De
  Ridder}, {de Souza}, {de Torres}, {del Peloso}, {del Pozo}, {Delbo},
  {Delgado}, {Delisle}, {Demouchy}, {Dharmawardena}, {Di Matteo}, {Diakite},
  {Diener}, {Distefano}, {Dolding}, {Edvardsson}, {Enke}, {Fabre}, {Fabrizio},
  {Faigler}, {Fedorets}, {Fernique}, {Fienga}, {Figueras}, {Fournier},
  {Fouron}, {Fragkoudi}, {Gai}, {Garcia-Gutierrez}, {Garcia-Reinaldos},
  {Garc{\'\i}a-Torres}, {Garofalo}, {Gavel}, {Gavras}, {Gerlach}, {Geyer},
  {Giacobbe}, {Gilmore}, {Girona}, {Giuffrida}, {Gomel}, {Gomez},
  {Gonz{\'a}lez-N{\'u}{\~n}ez}, {Gonz{\'a}lez-Santamar{\'\i}a},
  {Gonz{\'a}lez-Vidal}, {Granvik}, {Guillout}, {Guiraud},
  {Guti{\'e}rrez-S{\'a}nchez}, {Guy}, {Hatzidimitriou}, {Hauser}, {Haywood},
  {Helmer}, {Helmi}, {Sarmiento}, {Hidalgo}, {Hilger}, {H{\l}adczuk}, {Hobbs},
  {Holland}, {Huckle}, {Jardine}, {Jasniewicz}, {Jean-Antoine Piccolo},
  {Jim{\'e}nez-Arranz}, {Jorissen}, {Juaristi Campillo}, {Julbe}, {Karbevska},
  {Kervella}, {Khanna}, {Kontizas}, {Kordopatis}, {Korn}, {K{\'o}sp{\'a}l},
  {Kostrzewa-Rutkowska}, {Kruszy{\'n}ska}, {Kun}, {Laizeau}, {Lambert},
  {Lanza}, {Lasne}, {Le Campion}, {Lebreton}, {Lebzelter}, {Leccia}, {Leclerc},
  {Lecoeur-Taibi}, {Liao}, {Licata}, {Lindstr{\o}m}, {Lister}, {Livanou},
  {Lobel}, {Lorca}, {Loup}, {Madrero Pardo}, {Magdaleno Romeo}, {Managau},
  {Mann}, {Manteiga}, {Marchant}, {Marconi}, {Marcos}, {Marcos Santos},
  {Mar{\'\i}n Pina}, {Marinoni}, {Marocco}, {Marshall}, {Martin Polo},
  {Mart{\'\i}n-Fleitas}, {Marton}, {Mary}, {Masip}, {Massari},
  {Mastrobuono-Battisti}, {Mazeh}, {McMillan}, {Messina}, {Michalik}, {Millar},
  {Mints}, {Molina}, {Molinaro}, {Moln{\'a}r}, {Monari}, {Mongui{\'o}},
  {Montegriffo}, {Montero}, {Mor}, {Mora}, {Morbidelli}, {Morel}, {Morris},
  {Muraveva}, {Murphy}, {Musella}, {Nagy}, {Noval}, {Oca{\~n}a}, {Ogden},
  {Ordenovic}, {Osinde}, {Pagani}, {Pagano}, {Palaversa}, {Palicio},
  {Pallas-Quintela}, {Panahi}, {Payne-Wardenaar}, {Pe{\~n}alosa Esteller},
  {Penttil{\"a}}, {Pichon}, {Piersimoni}, {Pineau}, {Plachy}, {Plum}, {Poggio},
  {Pr{\v{s}}a}, {Pulone}, {Racero}, {Ragaini}, {Rainer}, {Raiteri}, {Rambaux},
  {Ramos}, {Ramos-Lerate}, {Re Fiorentin}, {Regibo}, {Richards}, {Rios Diaz},
  {Ripepi}, {Riva}, {Rix}, {Rixon}, {Robichon}, {Robin}, {Robin}, {Roelens},
  {Rogues}, {Rohrbasser}, {Romero-G{\'o}mez}, {Rowell}, {Royer}, {Ruz Mieres},
  {Rybicki}, {Sadowski}, {S{\'a}ez N{\'u}{\~n}ez}, {Sagrist{\`a} Sell{\'e}s},
  {Sahlmann}, {Salguero}, {Samaras}, {Sanchez Gimenez}, {Sanna},
  {Santove{\~n}a}, {Sarasso}, {Schultheis}, {Sciacca}, {Segol}, {Segovia},
  {S{\'e}gransan}, {Semeux}, {Shahaf}, {Siddiqui}, {Siebert}, {Siltala},
  {Silvelo}, {Slezak}, {Slezak}, {Smart}, {Snaith}, {Solano}, {Solitro},
  {Souami}, {Souchay}, {Spagna}, {Spina}, {Spoto}, {Steele},
  {Steidelm{\"u}ller}, {Stephenson}, {S{\"u}veges}, {Surdej}, {Szabados},
  {Szegedi-Elek}, {Taris}, {Taylor}, {Teixeira}, {Tolomei}, {Tonello}, {Torra},
  {Torra}, {Torralba Elipe}, {Trabucchi}, {Tsounis}, {Turon}, {Ulla}, {Unger},
  {Vaillant}, {van Dillen}, {van Reeven}, {Vanel}, {Vecchiato}, {Viala},
  {Vicente}, {Voutsinas}, {Weiler}, {Wevers}, {Wyrzykowski}, {Yoldas}, {Yvard},
  {Zhao}, {Zorec}, {Zucker}, \& {Zwitter}}]{2023gaiadr3}
{Gaia Collaboration}, {Vallenari}, A., {Brown}, A.~G.~A., {et~al.} 2023, \aap,
  674, A1, \dodoi{10.1051/0004-6361/202243940}

\bibitem[{{Gilmore} \& {Reid}(1983)}]{1983MNRAS.202.1025G}
{Gilmore}, G., \& {Reid}, N. 1983, \mnras, 202, 1025,
  \dodoi{10.1093/mnras/202.4.1025}

\bibitem[{{Gong} {et~al.}(2017){Gong}, {Henkel}, {Ott}, {Menten}, {Morris},
  {Keller}, {Claussen}, {Grasshoff}, \& {Mao}}]{2017ApJ...843...54G}
{Gong}, Y., {Henkel}, C., {Ott}, J., {et~al.} 2017, \apj, 843, 54,
  \dodoi{10.3847/1538-4357/aa7853}

\bibitem[{{Gray} {et~al.}(2009){Gray}, {Wittkowski}, {Scholz}, {Humphreys},
  {Ohnaka}, \& {Boboltz}}]{2009MNRAS.394...51G}
{Gray}, M.~D., {Wittkowski}, M., {Scholz}, M., {et~al.} 2009, \mnras, 394, 51,
  \dodoi{10.1111/j.1365-2966.2008.14237.x}

\bibitem[{{Habing}(1996)}]{Habing1996}
{Habing}, H.~J. 1996, \aapr, 7, 97, \dodoi{10.1007/PL00013287}

\bibitem[{{Hachenberg} {et~al.}(1973){Hachenberg}, {Grahl}, \&
  {Wielebinski}}]{1973IEEEP..61.1288H}
{Hachenberg}, O., {Grahl}, B.~H., \& {Wielebinski}, R. 1973, IEEE Proceedings,
  61, 1288

\bibitem[{{Helmi}(2020)}]{2020ARA&A..58..205H}
{Helmi}, A. 2020, \araa, 58, 205, \dodoi{10.1146/annurev-astro-032620-021917}

\bibitem[{{H{\"o}fner} \& {Olofsson}(2018)}]{2018A&ARv..26....1H}
{H{\"o}fner}, S., \& {Olofsson}, H. 2018, \aapr, 26, 1,
  \dodoi{10.1007/s00159-017-0106-5}

\bibitem[{{Iwanek} {et~al.}(2021){Iwanek}, {Soszy{\'n}ski}, \&
  {Koz{\l}owski}}]{2021ApJ...919...99I}
{Iwanek}, P., {Soszy{\'n}ski}, I., \& {Koz{\l}owski}, S. 2021, \apj, 919, 99,
  \dodoi{10.3847/1538-4357/ac10c5}

\bibitem[{{Iwanek} {et~al.}(2023){Iwanek}, {Poleski}, {Koz{\l}owski},
  {Soszy{\'n}ski}, {Pietrukowicz}, {Ban}, {Skowron}, {Mr{\'o}z}, {Wrona},
  {Udalski}, {Szyma{\'n}ski}, {Skowron}, {Ulaczyk}, {Gromadzki}, {Rybicki}, \&
  {Ratajczak}}]{2023ApJS..264...20I}
{Iwanek}, P., {Poleski}, R., {Koz{\l}owski}, S., {et~al.} 2023, \apjs, 264, 20,
  \dodoi{10.3847/1538-4365/acad7a}

\bibitem[{{Jeste} {et~al.}(2022){Jeste}, {Gong}, {Wong}, {Menten},
  {Kami{\'n}ski}, \& {Wyrowski}}]{2022A&A...666A..69J}
{Jeste}, M., {Gong}, Y., {Wong}, K.~T., {et~al.} 2022, \aap, 666, A69,
  \dodoi{10.1051/0004-6361/202243365}

\bibitem[{{Jiang} {et~al.}(1995){Jiang}, {Deguchi}, {Izumiura}, {Nakada}, \&
  {Yamamura}}]{1995PASJ...47..815J}
{Jiang}, B.~W., {Deguchi}, S., {Izumiura}, H., {Nakada}, Y., \& {Yamamura}, I.
  1995, \pasj, 47, 815

\bibitem[{{Juri{\'c}} {et~al.}(2008){Juri{\'c}}, {Ivezi{\'c}}, {Brooks},
  {Lupton}, {Schlegel}, {Finkbeiner}, {Padmanabhan}, {Bond}, {Sesar},
  {Rockosi}, {Knapp}, {Gunn}, {Sumi}, {Schneider}, {Barentine}, {Brewington},
  {Brinkmann}, {Fukugita}, {Harvanek}, {Kleinman}, {Krzesinski}, {Long},
  {Neilsen}, {Nitta}, {Snedden}, \& {York}}]{2008ApJ...673..864J}
{Juri{\'c}}, M., {Ivezi{\'c}}, {\v{Z}}., {Brooks}, A., {et~al.} 2008, \apj,
  673, 864, \dodoi{10.1086/523619}

\bibitem[{{Karachentsev} {et~al.}(2004){Karachentsev}, {Karachentseva},
  {Huchtmeier}, \& {Makarov}}]{2004AJ....127.2031K}
{Karachentsev}, I.~D., {Karachentseva}, V.~E., {Huchtmeier}, W.~K., \&
  {Makarov}, D.~I. 2004, \aj, 127, 2031, \dodoi{10.1086/382905}

\bibitem[{{Kim} {et~al.}(2014){Kim}, {Cho}, \& {Kim}}]{2014AJ....147...22K}
{Kim}, J., {Cho}, S.-H., \& {Kim}, S.~J. 2014, \aj, 147, 22,
  \dodoi{10.1088/0004-6256/147/1/22}

\bibitem[{{Klein} {et~al.}(2012){Klein}, {Hochg{\"u}rtel}, {Kr{\"a}mer},
  {Bell}, {Meyer}, \& {G{\"u}sten}}]{2012A&A...542L...3K}
{Klein}, B., {Hochg{\"u}rtel}, S., {Kr{\"a}mer}, I., {et~al.} 2012, \aap, 542,
  L3, \dodoi{10.1051/0004-6361/201218864}

\bibitem[{{Kwon} \& {Suh}(2012)}]{2012JKAS...45..139K}
{Kwon}, Y.-J., \& {Suh}, K.-W. 2012, \jkas, 45, 139,
  \dodoi{10.5303/JKAS.2012.45.6.139}

\bibitem[{{Le Squeren} {et~al.}(1992){Le Squeren}, {Sivagnanam}, {Dennefeld},
  \& {David}}]{1992A&A...254..133L}
{Le Squeren}, A.~M., {Sivagnanam}, P., {Dennefeld}, M., \& {David}, P. 1992,
  \aap, 254, 133

\bibitem[{{Lewis} {et~al.}(2020){Lewis}, {Pihlstr{\"o}m}, {Sjouwerman},
  {Stroh}, {Morris}, \& {BAaDE Collaboration}}]{2020ApJ...892...52L}
{Lewis}, M.~O., {Pihlstr{\"o}m}, Y.~M., {Sjouwerman}, L.~O., {et~al.} 2020,
  \apj, 892, 52, \dodoi{10.3847/1538-4357/ab7920}

\bibitem[{{Li} {et~al.}(2019){Li}, {FELLOW}, {Liu}, {Xue}, {Zhong}, {Weiss},
  {Carlin}, {Tian}, \& {FELLOW}}]{2019ApJ...874..138L}
{Li}, J., {FELLOW}, L., {Liu}, C., {et~al.} 2019, \apj, 874, 138,
  \dodoi{10.3847/1538-4357/ab09ef}

\bibitem[{{Lindegren} {et~al.}(2021){Lindegren}, {Bastian}, {Biermann},
  {Bombrun}, {de Torres}, {Gerlach}, {Geyer}, {Hern{\'a}ndez}, {Hilger},
  {Hobbs}, {Klioner}, {Lammers}, {McMillan}, {Ramos-Lerate},
  {Steidelm{\"u}ller}, {Stephenson}, \& {van Leeuwen}}]{2021A+A...649A...4L}
{Lindegren}, L., {Bastian}, U., {Biermann}, M., {et~al.} 2021, \aap, 649, A4,
  \dodoi{10.1051/0004-6361/202039653}

\bibitem[{{Lindqvist} {et~al.}(1992){Lindqvist}, {Winnberg}, {Habing}, \&
  {Matthews}}]{1992A&AS...92...43L}
{Lindqvist}, M., {Winnberg}, A., {Habing}, H.~J., \& {Matthews}, H.~E. 1992,
  \aaps, 92, 43

\bibitem[{{Ma{\'\i}z Apell{\'a}niz}(2022)}]{2022A+A...657A.130M}
{Ma{\'\i}z Apell{\'a}niz}, J. 2022, \aap, 657, A130,
  \dodoi{10.1051/0004-6361/202142365}

\bibitem[{{Mauron} {et~al.}(2019){Mauron}, {Maurin}, \& {Kendall}}]{Mauron2019}
{Mauron}, N., {Maurin}, L.~P.~A., \& {Kendall}, T.~R. 2019, \aap, 626, A112,
  \dodoi{10.1051/0004-6361/201834089}

\bibitem[{{Menten} {et~al.}(2018){Menten}, {Wyrowski}, {Keller}, \&
  {Kami{\'n}ski}}]{2018A&A...613A..49M}
{Menten}, K.~M., {Wyrowski}, F., {Keller}, D., \& {Kami{\'n}ski}, T. 2018,
  \aap, 613, A49, \dodoi{10.1051/0004-6361/201732296}

\bibitem[{{Messineo} {et~al.}(2002){Messineo}, {Habing}, {Sjouwerman}, {Omont},
  \& {Menten}}]{2002A&A...393..115M}
{Messineo}, M., {Habing}, H.~J., {Sjouwerman}, L.~O., {Omont}, A., \& {Menten},
  K.~M. 2002, \aap, 393, 115, \dodoi{10.1051/0004-6361:20021017}

\bibitem[{{Messineo} {et~al.}(2018){Messineo}, {Habing}, {Sjouwerman}, {Omont},
  \& {Menten}}]{2018A&A...619A..35M}
---. 2018, \aap, 619, A35, \dodoi{10.1051/0004-6361/201730717}

\bibitem[{{M{\"u}ller} {et~al.}(2005){M{\"u}ller}, {Schl{\"o}der}, {Stutzki},
  \& {Winnewisser}}]{2005JMoSt.742..215M}
{M{\"u}ller}, H. S.~P., {Schl{\"o}der}, F., {Stutzki}, J., \& {Winnewisser}, G.
  2005, \jmost, 742, 215, \dodoi{10.1016/j.molstruc.2005.01.027}

\bibitem[{{Nissen} \& {Schuster}(2010)}]{2010A&A...511L..10N}
{Nissen}, P.~E., \& {Schuster}, W.~J. 2010, \aap, 511, L10,
  \dodoi{10.1051/0004-6361/200913877}

\bibitem[{{Palaversa} {et~al.}(2013){Palaversa}, {Ivezi{\'c}}, {Eyer},
  {Ru{\v{z}}djak}, {Sudar}, {Galin}, {Kroflin}, {Mesari{\'c}}, {Munk},
  {Vrbanec}, {Bo{\v{z}}i{\'c}}, {Loebman}, {Sesar}, {Rimoldini}, {Hunt-Walker},
  {VanderPlas}, {Westman}, {Stuart}, {Becker}, {Srdo{\v{c}}}, {Wozniak}, \&
  {Oluseyi}}]{2013AJ....146..101P}
{Palaversa}, L., {Ivezi{\'c}}, {\v{Z}}., {Eyer}, L., {et~al.} 2013, \aj, 146,
  101, \dodoi{10.1088/0004-6256/146/4/101}

\bibitem[{{Pardo} {et~al.}(2004){Pardo}, {Alcolea}, {Bujarrabal}, {Colomer},
  {del Romero}, \& {de Vicente}}]{2004A&A...424..145P}
{Pardo}, J.~R., {Alcolea}, J., {Bujarrabal}, V., {et~al.} 2004, \aap, 424, 145,
  \dodoi{10.1051/0004-6361:20040309}

\bibitem[{{Pety}(2005)}]{2005sf2a.conf..721P}
{Pety}, J. 2005, in SF2A-2005: Semaine de l'Astrophysique Francaise, ed.
  F.~{Casoli}, T.~{Contini}, J.~M. {Hameury}, \& L.~{Pagani} (Les Ulis:
  EdP-Sciences), 721

\bibitem[{{Reid} \& {Dickinson}(1976)}]{Reid1976}
{Reid}, M.~J., \& {Dickinson}, D.~F. 1976, \apj, 209, 505,
  \dodoi{10.1086/154745}

\bibitem[{{Reid} \& {Honma}(2014)}]{2014ARA&A..52..339R}
{Reid}, M.~J., \& {Honma}, M. 2014, \araa, 52, 339,
  \dodoi{10.1146/annurev-astro-081913-040006}

\bibitem[{{Reid} {et~al.}(2009){Reid}, {Menten}, {Zheng}, {Brunthaler},
  {Moscadelli}, {Xu}, {Zhang}, {Sato}, {Honma}, {Hirota}, {Hachisuka}, {Choi},
  {Moellenbrock}, \& {Bartkiewicz}}]{2009ApJ...700..137R}
{Reid}, M.~J., {Menten}, K.~M., {Zheng}, X.~W., {et~al.} 2009, \apj, 700, 137,
  \dodoi{10.1088/0004-637X/700/1/137}

\bibitem[{{Reid} {et~al.}(2019){Reid}, {Menten}, {Brunthaler}, {Zheng}, {Dame},
  {Xu}, {Li}, {Sakai}, {Wu}, {Immer}, {Zhang}, {Sanna}, {Moscadelli}, {Rygl},
  {Bartkiewicz}, {Hu}, {Quiroga-Nu{\~n}ez}, \& {van
  Langevelde}}]{2019ApJ...885..131R}
{Reid}, M.~J., {Menten}, K.~M., {Brunthaler}, A., {et~al.} 2019, \apj, 885,
  131, \dodoi{10.3847/1538-4357/ab4a11}

\bibitem[{{Scholz} \& {Wood}(2000)}]{Scholz2000}
{Scholz}, M., \& {Wood}, P.~R. 2000, \aap, 362, 1065

\bibitem[{{Sevenster}(1999)}]{1999MNRAS.310..629S}
{Sevenster}, M.~N. 1999, \mnras, 310, 629,
  \dodoi{10.1046/j.1365-8711.1999.02957.x}

\bibitem[{{Sevenster} {et~al.}(1997){Sevenster}, {Chapman}, {Habing},
  {Killeen}, \& {Lindqvist}}]{1997A&AS..122...79S}
{Sevenster}, M.~N., {Chapman}, J.~M., {Habing}, H.~J., {Killeen}, N.~E.~B., \&
  {Lindqvist}, M. 1997, \aaps, 122, 79, \dodoi{10.1051/aas:1997294}

\bibitem[{{Sjouwerman} {et~al.}(1998){Sjouwerman}, {van Langevelde},
  {Winnberg}, \& {Habing}}]{1998A&AS..128...35S}
{Sjouwerman}, L.~O., {van Langevelde}, H.~J., {Winnberg}, A., \& {Habing},
  H.~J. 1998, \aaps, 128, 35, \dodoi{10.1051/aas:1998127}

\bibitem[{{Skowron} {et~al.}(2019){Skowron}, {Skowron}, {Mr{\'o}z}, {Udalski},
  {Pietrukowicz}, {Soszy{\'n}ski}, {Szyma{\'n}ski}, {Poleski}, {Koz{\l}owski},
  {Ulaczyk}, {Rybicki}, \& {Iwanek}}]{2019Sci...365..478S}
{Skowron}, D.~M., {Skowron}, J., {Mr{\'o}z}, P., {et~al.} 2019, Science, 365,
  478, \dodoi{10.1126/science.aau3181}

\bibitem[{{Sparke} \& {Gallagher}(2007)}]{2007gitu.book.....S}
{Sparke}, L.~S., \& {Gallagher}, J.~S., I. 2007, {Galaxies in the Universe: An
  Introduction (2nd ed.; Cambridge: Cambridge Univ. Press)}

\bibitem[{{Steinmetz} {et~al.}(2020){Steinmetz}, {Matijevi{\v{c}}}, {Enke},
  {Zwitter}, {Guiglion}, {McMillan}, {Kordopatis}, {Valentini}, {Chiappini},
  {Casagrande}, {Wojno}, {Anguiano}, {Bienaym{\'e}}, {Bijaoui}, {Binney},
  {Burton}, {Cass}, {de Laverny}, {Fiegert}, {Freeman}, {Fulbright}, {Gibson},
  {Gilmore}, {Grebel}, {Helmi}, {Kunder}, {Munari}, {Navarro}, {Parker},
  {Ruchti}, {Recio-Blanco}, {Reid}, {Seabroke}, {Siviero}, {Siebert}, {Stupar},
  {Watson}, {Williams}, {Wyse}, {Anders}, {Antoja}, {Birko}, {Bland-Hawthorn},
  {Bossini}, {Garc{\'\i}a}, {Carrillo}, {Chaplin}, {Elsworth}, {Famaey},
  {Gerhard}, {Jofre}, {Just}, {Mathur}, {Miglio}, {Minchev}, {Monari},
  {Mosser}, {Ritter}, {Rodrigues}, {Scholz}, {Sharma}, {Sysoliatina}, \& {RAVE
  Collaboration}}]{2020AJ....160...82S}
{Steinmetz}, M., {Matijevi{\v{c}}}, G., {Enke}, H., {et~al.} 2020, \aj, 160,
  82, \dodoi{10.3847/1538-3881/ab9ab9}

\bibitem[{{Stroh} {et~al.}(2018){Stroh}, {Pihlstr{\"o}m}, {Sjouwerman},
  {Claussen}, {Morris}, \& {Rich}}]{2018ApJ...862..153S}
{Stroh}, M.~C., {Pihlstr{\"o}m}, Y.~M., {Sjouwerman}, L.~O., {et~al.} 2018,
  \apj, 862, 153, \dodoi{10.3847/1538-4357/aaccf3}

\bibitem[{{Stroh} {et~al.}(2019){Stroh}, {Pihlstr{\"o}m}, {Sjouwerman},
  {Lewis}, {Claussen}, {Morris}, \& {Rich}}]{2019ApJS..244...25S}
---. 2019, \apjs, 244, 25, \dodoi{10.3847/1538-4365/ab3c35}

\bibitem[{{Suh}(2021)}]{2021ApJS..256...43S}
{Suh}, K.-W. 2021, \apjs, 256, 43, \dodoi{10.3847/1538-4365/ac1274}

\bibitem[{{Sun} {et~al.}(2022){Sun}, {Zhang}, {Reid}, {Xu}, {Wen}, {Zhang}, \&
  {Zheng}}]{2022ApJ...931...74S}
{Sun}, Y., {Zhang}, B., {Reid}, M.~J., {et~al.} 2022, \apj, 931, 74,
  \dodoi{10.3847/1538-4357/ac69e0}

\bibitem[{{te Lintel Hekkert} {et~al.}(1991){te Lintel Hekkert}, {Caswell},
  {Habing}, {Haynes}, {Haynes}, \& {Norris}}]{1991A&AS...90..327T}
{te Lintel Hekkert}, P., {Caswell}, J.~L., {Habing}, H.~J., {et~al.} 1991,
  \aaps, 90, 327

\bibitem[{{Trabucchi}(2023)}]{2023arXiv230617758T}
{Trabucchi}, M. 2023, arXiv e-prints, arXiv:2306.17758,
  \dodoi{10.48550/arXiv.2306.17758}

\bibitem[{{Trapp} {et~al.}(2018){Trapp}, {Rich}, {Morris}, {Sjouwerman},
  {Pihlstr{\"o}m}, {Claussen}, \& {Stroh}}]{2018ApJ...861...75T}
{Trapp}, A.~C., {Rich}, R.~M., {Morris}, M.~R., {et~al.} 2018, \apj, 861, 75,
  \dodoi{10.3847/1538-4357/aac382}

\bibitem[{{Venn} {et~al.}(2004){Venn}, {Irwin}, {Shetrone}, {Tout}, {Hill}, \&
  {Tolstoy}}]{2004AJ....128.1177V}
{Venn}, K.~A., {Irwin}, M., {Shetrone}, M.~D., {et~al.} 2004, \aj, 128, 1177,
  \dodoi{10.1086/422734}

\bibitem[{{Wang} {et~al.}(2017){Wang}, {Yu}, {Jiang}, {Zhao}, {Sun}, {Li},
  {Zhong}, {Dong}, {Michael}, {Xia}, {Zuo}, {Gou}, {Guo}, {Lu}, {Liu}, {Fan},
  {Jiang}, \& {Qian}}]{2017AcASn..58...37W}
{Wang}, J.~Q., {Yu}, L.~F., {Jiang}, Y.~B., {et~al.} 2017, \acasn, 58, 37

\bibitem[{{Wielebinski} {et~al.}(2011){Wielebinski}, {Junkes}, \&
  {Grahl}}]{2011JAHH...14....3W}
{Wielebinski}, R., {Junkes}, N., \& {Grahl}, B.~H. 2011, Journal of
  Astronomical History and Heritage, 14, 3

\bibitem[{{Wu} {et~al.}(2022){Wu}, {Zhang}, {Li}, \&
  {Zheng}}]{2022MNRAS.516.1881W}
{Wu}, Y., {Zhang}, B., {Li}, J., \& {Zheng}, X.-W. 2022, \mnras, 516, 1881,
  \dodoi{10.1093/mnras/stac1971}

\bibitem[{{Wu} {et~al.}(2018){Wu}, {Matsunaga}, {Burns}, \&
  {Zhang}}]{2018MNRAS.473.3325W}
{Wu}, Y.~W., {Matsunaga}, N., {Burns}, R.~A., \& {Zhang}, B. 2018, \mnras, 473,
  3325, \dodoi{10.1093/mnras/stx2450}

\bibitem[{{Yan} {et~al.}(2015){Yan}, {Shen}, {Wu}, {Manchester}, {Weltevrede},
  {Wu}, {Zhao}, {Yuan}, {Lee}, {Fan}, {Hong}, {Jiang}, {Li}, {Liang}, {Ling},
  {Liu}, {Qian}, {Zhang}, {Zhong}, \& {Ye}}]{2015ApJ...814....5Y}
{Yan}, Z., {Shen}, Z.-Q., {Wu}, X.-J., {et~al.} 2015, \apj, 814, 5,
  \dodoi{10.1088/0004-637X/814/1/5}

\bibitem[{{Young}(1995)}]{1995ApJ...445..872Y}
{Young}, K. 1995, \apj, 445, 872, \dodoi{10.1086/175747}

\bibitem[{{Zhang} {et~al.}(2012){Zhang}, {Reid}, {Menten}, \&
  {Zheng}}]{2012ApJ...744...23Z}
{Zhang}, B., {Reid}, M.~J., {Menten}, K.~M., \& {Zheng}, X.~W. 2012, \apj, 744,
  23, \dodoi{10.1088/0004-637X/744/1/23}

\bibitem[{{Zhong} {et~al.}(2018){Zhong}, {Dong}, {Gou}, {Yu}, {Wang}, {Xia},
  {Jiang}, {Liu}, {Zhang}, {Shi}, {Yin}, {Shi}, {Liu}, \&
  {Shen}}]{2018RAA....18...44Z}
{Zhong}, W.-Y., {Dong}, J., {Gou}, W., {et~al.} 2018, \raa, 18, 044,
  \dodoi{10.1088/1674-4527/18/4/44}

\end{thebibliography}
\bibliographystyle{aasjournal}



\end{document}